\documentclass{aa}  

\usepackage{graphicx}
\usepackage{xcolor}
\usepackage{subcaption}
\usepackage{multirow}
\usepackage{tabularx}
\usepackage{upgreek}
\usepackage{txfonts}
\usepackage{hyperref}
\hypersetup{
    colorlinks=true,
    linkcolor=blue,
    filecolor=blue,
    urlcolor=blue,
    citecolor=blue,
}

\bibpunct{(}{)}{;}{a}{}{,} 

\newcommand{\dd}{\ensuremath{\mathrm{d}}}

\newcommand{\simno}{Sim.~No.}

\begin{document} 

   \title{Magnetic field and kinetic helicity evolution in simulations of interacting disk galaxies}
   \titlerunning{Magnetic field evolution in interacting disk galaxy simulations}

   \author{S. Selg
          \inst{1}
          \and
          W. Schmidt
          \inst{1}
          }
   \authorrunning{Selg and Schmidt}

   \institute{Hamburger Sternwarte, Universität Hamburg, Gojenbergsweg 112,
             D-21029 Hamburg, Germany\\
             \email{wolfram.schmidt@uni-hamburg.de}
             }

   \date{Manuscript}

 
  \abstract
   {There are indications that the magnetic field evolution in galaxies is influenced by tidal interactions and mergers between galaxies.}
   {We carried out a parameter study of interacting disk galaxies with impact parameters ranging from central collisions to weakly interacting scenarios. The orientations of the disks were also varied. In particular, we investigated how magnetic field amplification depends on these parameters.}
   {We used magnetohydrodynamics for gas disks in combination with live dark-matter halos in adaptive mesh refinement simulations. The disks were initialized using a setup for isolated disks in hydrostatic equilibrium. Since we focused on the impact of tidal forces on magnetic field evolution, adiabatic physics was applied. Small-scale filtering of the velocity and magnetic field allowed us to estimate the turbulent electromotive force (EMF) and kinetic helicity.}
   {Time series of the average magnetic field in central and outer disk regions show pronounced peaks during close encounters and mergers. This agrees with observed magnetic fields at different interaction stages. The central field strength exceeds $10\;\upmu$G (corresponding to an amplification factor of 2 to 3) for small impact parameters. As the disks are increasingly disrupted and turbulence is produced by tidal forces, the small-scale EMF reaches a significant fraction of the total EMF. The small-scale kinetic helicity is initially antisymmetric across the disk plane. Though its evolution is sensitive to both the impact parameter and inclinations of the rotation axes with respect to the relative motion of the disks, antisymmetry is generally broken through interactions and the merger remnant loses most of the initial helicity. The EMF and the magnetic field also decay rapidly after coalescence.}  
   {The strong amplification during close encounters of the interacting galaxies is mostly driven by helical flows and a mean-field dynamo. The small-scale dynamo contributes significantly in post-interaction phases. However, the amplification of the magnetic field cannot be sustained.}


   \maketitle
%

\section{Introduction}
\label{sec:intro}

Observations show that galaxy interactions and mergers play a significant role during the evolution of galaxies \citep{Struck99, Smith10}. In addition to increasing the star formation rate \citep{Kennicutt98,Georgakakis00,Rossa07}, there are indications that mergers have an impact on the magnetic field strength in galaxies \citep{Condon02,Chyzy04,Drzazga11}, suggesting an amplification of the field due to dynamo action. 
While the $\alpha$-$\Omega$ dynamo induced by the differential rotation of disk galaxies is thought to play an important role in the generation of galactic magnetic fields from a weak primordial field \citep{Brandenburg05,Park13,Steinwandel2020,Wissing2023}, the $\alpha^2$ dynamo (caused by inhomogeneous shear flows with kinetic helicity) and the turbulent small-scale dynamo (induced by non-helical, isotropic turbulence on small scales) are expected to contribute significantly to the field amplification in merging galaxies \citep{Arshakian2009,Schober2013,Bhat14,Moss2014}. In contrast to isolated disk galaxies, where turbulence is assumed to be driven by supernova explosions \citep{Hopkins13} or gravitational instabilities \citep{Krumholz16}, shear and tidal effects can produce additional turbulence during galaxy interactions \citep{Renaud14,Whittingham21,Jimenez2023}.

The interaction process has been extensively studied in idealized numerical simulations without cosmological initial conditions. The basic idea is to embed disks composed of gas and possibly stars in spherical dark-matter halos and to set them on a collision course. An original study of such systems using a simple particle-based model was presented by \cite{Toomre72} five decades ago. More recently, smoothed particle hydrodynamic simulations \citep{Mihos96,DiMatteo08,Blumenthal18} and semi-Lagrangian simulations \citep{Hayward14,Moreno19} focused on the dynamics of the interstellar medium and interaction-induced star formation. Merger simulations with magnetic fields were first performed by \citet{Kotarba10}, using a magnetohydrodynamic (MHD) extension of smoothed particle hydrodynamics \citep{Dolag09}. Their simulation of the Antennae galaxies suggests that mergers play a significant role in the amplification of galactic magnetic fields. Magnetic field amplification during minor mergers, where the secondary galaxy has only a small fraction of the mass of the primary galaxy, was investigated with a similar approach by \citet{Geng12} and in cosmological simulations with the semi-Lagrangian Arepo code by \citet{Pakmor2014}. Morphological changes due to mergers were studied for a sample of galaxies from the Horizon-AGN cosmological simulation \citep{Martin2018}. Moreover, the role of magnetic fields and the small-scale dynamo was analyzed in great detail in zoom-in simulations of major mergers from the Illustris simulation by \citet{Whittingham21}. However, in contrast to simulating pairs of galaxies, the resolution of cosmological simulations is limited and interaction parameters cannot be varied systematically. 

While Lagrangian methods undoubtedly have merits for the treatment of moving galaxies, \citet{Rodenbeck2016} demonstrate that merger simulations are feasible in a greatly simplified setup with purely gaseous and adiabatic disks without dark-matter halos using the MHD adaptive mesh refinement (AMR) code Enzo \citep{Enzo2014}. They applied the method from \citet{Wang10} to initialize an isolated equilibrium disk in a static external potential that represents the gravity of dark matter with additional support from magnetic pressure. However, they excluded the external potential because it is not applicable to moving disks in a galaxy merger scenario, although the contribution of dark matter is dominant. Even so, this approach avoids the difficulties of implementing MHD into particle-based codes \citep{Dolag09} and, owing to a well-defined resolution scale, offers the advantage of better controllability of the small-scale physics. For realistic simulations of interacting galaxies, however, N-body dynamics is essential. Hydrodynamic AMR simulations of a major merger similar to the Antennae system with live N-body halos were performed by \citet{Teyssier10} and \citet{Renaud2015}. At spatial resolutions in the parsec range, they were able to analyze the production of turbulence in the interstellar medium due to tidal interactions and the compression-induced enhancement of star formation. 

In this paper, we combine the approaches of \citet{Teyssier10} and \citet{Rodenbeck2016}. To generate initial conditions for spherical dark-matter halos, we follow \citet{Drakos17}. The gaseous disks are based on \citet{Wang10}. In essence, the gas density and rotation velocity of a disk in hydrostatic equilibrium are determined iteratively from the self-gravity of the gas and the smoothed potential of the dark-matter halo. The advantage of this method of initializing disks is that the effects of the initial relaxation are reduced and changes in the disk structure are mainly due to physical effects, such as tidal forces and fragmentation induced by cooling and gravity. The applied disk and halo models are described in more detail in the following section. In Sect.~\ref{sec:numerics}, the numerical setup is outlined and an overview of our simulation suite is provided. In this study we focus on magnetic field amplification due to tidal interactions between the disks and their halos. While radiative cooling and stellar feedback are essential ingredients for a realistic disk structure, the findings of \citet{Ntormousi2020} indicate that these processes are of minor importance for the generation of galactic magnetic fields. Since nonadiabatic physics increases computational requirements substantially for the targeted spatial resolution of about $10\;$pc, we use adiabatic gas dynamics in this work. The analysis of the simulation is presented in Sect.~\ref{sec:results}: After a phenomenological discussion of representative examples, the average magnetic fields in central and outer disk regions from different simulations and subsequent encounters of the interacting galaxies in each simulation are compared. Moreover, the question of whether magnetic field amplification during interaction phases can be related to dynamo action on the smallest numerically resolved length scales is investigated. In the conclusions, we sum up and discuss our results. 


\section{Disk and halo models}
\label{sec:models}

To initialize the gaseous disks, we applied the potential method of \citet{Wang10} for a disk with exponential surface density:
\begin{equation}
    \label{eq:exp_surf}
    \Sigma(r) := \int_\infty^\infty \rho(r,z)\dd z = \Sigma_0\exp(-r/r_\mathrm{d})\,,
\end{equation}
where $r$ and $z$ are, respectively, the radius and height in a cylindrical coordinate system with the disk midplane at $z=0$. The surface density at the center is given by $\Sigma_0 = M_\mathrm{d}/2\uppi r_\mathrm{d}^2$ for a disk of total mass $M_\mathrm{d}$ and scale length $r_\mathrm{d}$. In many isolated disk simulations, an exponential form is assumed for the volume density $\rho(r,z)$ of the gas. However, as argued by \citet{Wang10}, this implies a ring in the surface density. 

The gas density is obtained by integrating the equation of hydrostatic equilibrium in vertical direction. For an initially isothermal disk composed of an ideal gas and a magnetic field with a constant ratio of thermal to magnetic pressure, $\beta=p/p_{\mathrm{m}}$, we have the total pressure
\begin{equation}
    p + p_{\mathrm{m}} =
    (\gamma-1)\rho e + \frac{B^2}{8\uppi}
    = \left(1 + \epsilon_\mathrm{m}\right)\rho c_0^2\,,
\end{equation}
where $e$ is the specific internal energy of gas with adiabatic index $\gamma$,  $\epsilon_\mathrm{m}=1/\beta$ vanishes in the hydrodynamic limit, and the isothermal speed of sound is defined by
\begin{equation}
    c_0^2 := \frac{p}{\rho} = \frac{k_\mathrm{B}T}{\mu m_\mathrm{H}}
,\end{equation}
with Boltzmann's constant $k_\mathrm{B}$, the gas temperature $T$, the mean atomic weight $\mu$, and the hydrogen mass $m_\mathrm{H}$. In this case, the gas density can be expressed as (see \citealt{Wang10})
\begin{equation}
    \label{eq:rho_disk}
    \rho(r,z) = 
    \rho_{0}(r)\exp{\left(-\frac{\Phi_z(r,z)}{(1 + \epsilon_\mathrm{m})c_0^2}\right)}\,,
\end{equation}
where $\Phi_z(r,z)=\Phi(r,z)-\Phi(r,0)$ is the vertical potential difference of the total gravitational potential 
\begin{equation}
    \Phi(r,z) = \Phi_{\mathrm{dm}}(r,z) + \Phi_{\mathrm{s}}(r,z) + \Phi_{\mathrm{g}}(r,z)
\end{equation}
at height $z$ relative to the midplane. The dark matter potential $\Phi_{\mathrm{dm}}(r,z)$ follows from the halo model (see below) and the stellar potential $\Phi_{\mathrm{s}}(r,z)$ is zero in this work.

From the Poisson equation, it follows that the gravitational potential of the gas, $\Phi_{\mathrm{g}}(r,z)$, is given by
\begin{equation}
    \label{eq:pot_gas}
    \frac{\dd^2\Phi_{\mathrm{g}}(r,z)}{\dd z^2} = 
    4\uppi G\rho_0(r)\exp{\left(-\frac{\Phi_z(r,z)}{(1 + \epsilon_\mathrm{m})c_0^2}\right)}\,.
\end{equation}
Equation~(\ref{eq:exp_surf}) is satisfied if the midplane density $\rho_0(r)=\rho(r,0)$ is given by
\begin{equation}
    \label{eq:rho_midplane}
    \rho_0(r) = \frac{\Sigma(r)}{
        \int_\infty^\infty \exp{\left(-\Phi_z(r,z)/
            [(1 + \epsilon_\mathrm{m})c_0^2]\right)}\dd z}\,. 
\end{equation}
The integral in the denominator can be interpreted as a disk scale height that varies with radius and is determined by the model. Equations~(\ref{eq:pot_gas}) and (\ref{eq:rho_midplane}) cannot be solved analytically. A numerical solution is readily obtained by using an iterative scheme in combination with a numerical solver for the differential equation (see Sect.~\ref{sec:numerics}). An overview of all model parameters is provided in Table~\ref{tab:model}.

\begin{table}
\caption{Overview of the initial disk and halo model parameters.}
\label{tb:parameters}
\centering
\begin{tabular}{llcc}
  \hline\hline
  Comp. & \multicolumn{3}{c}{Parameter} \\ \hline
  \multirow{5}*{Disk} 
    & gas mass & $M_\mathrm{d}$ & $10^{10}\;\mathrm{M}_{\odot}$\\
    & scale length & $r_\mathrm{d}$ & $3.5\;\mathrm{kpc}$ \\
    & temperature & $T$ & $10^4\;\mathrm{K}$ \\ 
    & molecular weight & $\mu$ & $1.0$ \\ 
    & inverse plasma beta & $\epsilon_{\mathrm{m}}$ & $0.05$ \\
  \hline
  \multirow{3}*{Halo} 
    & asymptotic mass & $M_{\mathrm{dm}}$ & $1.3248\times 10^{12}\;\mathrm{M}_{\odot}$ \\
    & virial mass & $M_{\mathrm{200}}$ & $1\times 10^{12}\;\mathrm{M}_{\odot}$ \\
    & scale length & $a$ & $32\;\mathrm{kpc}$ \\
    & virial radius & $r_{200}$ & $210\;\mathrm{kpc}$ \\ 
  \hline
\end{tabular}
\label{tab:model}
\end{table}

For the dark-matter halo, we assumed the radial density profile \citep{Hernquist90}
\begin{equation}
    \rho_{\mathrm{dm}}(r) = \frac{M_{\mathrm{dm}}}{2\uppi}\frac{a}{r(r+a)^{3}}\,,
\end{equation}
where $a$ is the scale length of the halo and $M_{\mathrm{dm}}$ the total cumulative mass for $r\rightarrow\infty$. Here, $r$ is a spherical radial coordinate. As proposed by \cite{Springel05b}, the scale length of the Hernquist halo is chosen such that it contains the same mass $M_{200}$ within the virial radius $r_{200}$ as a Navarro–Frenk–White halo of given scale length $r_{\mathrm{s}}$ and concentration parameter c=$r_{200}/r_{\mathrm{s}}$:
\begin{equation}
    a = r_{\mathrm{s}}\sqrt{2[\ln{(1+c)} - c/(1+c)]}\,.
\end{equation}
The analytic form of the gravitational potential is
\begin{equation}
    \Phi_{\mathrm{dm}}(r) = -\frac{G M_{\mathrm{dm}}}{r + a}\,.
\end{equation}
We used the method developed by \citet{Drakos17} to compute random initial positions and velocities of the particles. The algorithm
populates the phase space of an N-body system based on the distribution function of a halo in virial equilibrium.

The rotation curve of the disk is given by \citep{Wang10,Rodenbeck2016} 
\begin{equation}
    v_{\mathrm{rot}}^{2}(r) = v_{\mathrm{dm}}^2(r) + v_{\mathrm{thin}}^{2}(r) + 
    (1+\epsilon_{\mathrm{m}})c_0^2\frac{\partial\ln{\rho}}{\partial\ln{r}}\bigg\vert_{\mathrm{z=0}},\label{eq:rotation01}
\end{equation}
where
\begin{equation}
    \label{eq:v_dm}
    v_{\mathrm{dm}}(r) = \frac{\sqrt{G M_{\mathrm{dm}} r}}{r + a}\,
\end{equation}
for a Hernquist halo, $v_{\mathrm{thin}}(r)$ is the contribution from an infinitesimally thin disk with exponential surface density defined by Eq.~(\ref{eq:exp_surf}) \citep[for details, see][]{Binney08}, and the last term is a correction due to the pressure gradient of the gas following from Eq.~(\ref{eq:rho_disk}).

Finally, the initial magnetic field is assumed to be toroidal with constant $\beta$ inside the disk \citep{Rodenbeck2016}:
\begin{equation}
    B_\phi = \sqrt{\frac{8\uppi\rho c_0^2}{\beta}} = \sqrt{8\uppi\epsilon_{\mathrm{m}}\rho c_0^2}\,,
\end{equation}
and $B_r = B_z = 0$ in the disk's cylindrical coordinate system. The disk interior is defined by $\rho \ge \rho_\mathrm{min}$, where $\rho_\mathrm{min}=6.768\times 10^{-30}\;\mathrm{g}\;\mathrm{cm}^{-3}$ is the constant density of the ambient medium. While we accounted for magnetic pressure in the computation of the equilibrium density and rotation curve, a toroidal magnetic field gives rise to an additional curvature force that is directed radially inward. In the case of a constant plasma beta, this force gives rise to a constant negative contribution to the rotation curve. We neglected this contribution because it is lower by several orders of magnitude than the maximum rotation speed for the chosen disk parameters.

The combination of a rotating disk in hydrostatic equilibrium with an N-body halo and a constant background gas density introduces several effects impeding the stability of the disk:
\begin{itemize}
\item Adiabatic disks have relatively large scale heights, particularly in the outer regions, while the model of \citet{Wang10} is based on a thin-disk approximation. Near the center, the scale height is small, but the disk structure cannot be mapped accurately to the grid due to the limited numerical resolution. In addition, the rotation velocity is independent of the height $z$ as long as the temperature remains constant, but adiabatically evolving perturbations give rise to $z$-dependent deviations from the initial rotation curve.
\item The density of the gaseous disk given by Eq.~(\ref{eq:rho_disk}) decreases rapidly with the scale height. To avoid unphysically low densities and density jumps at the interface between disk and ambient medium, the density floor $\rho_\mathrm{min}$ mentioned above is applied when setting up the disk. Since a constant density is not consistent with the equilibrium solution, gas is subsequently falling toward the disk. This could be alleviated by modeling a circumgalactic medium (see, for example, \citealt{Steinwandel2019}). However, the total mass in the ambient medium is very small compared to the disk mass.
\item Since the gravitational potential is determined by discrete particles, changes in the particle positions entail fluctuations of the potential, which in turn affect the gas.
\item Although the gaseous disk is computed from the joint potential of halo and disk, the distribution of dark matter particles is computed assuming the analytical potential of the halo and neglecting the gravity of the disk.
\end{itemize}
As a consequence, perturbations of the initial equilibrium disk are unavoidable. For the most part, the disk does not undergo abrupt changes, but gradually (over a few hundred$\;$megayears) evolves into a disk that exhibits a spiral-like structure.


\section{Numerical methods and overview of simulations}
\label{sec:numerics}

We used a modified version of the publicly available cosmological AMR code Enzo \citep{Enzo2014,Enzo2019}.\footnote{See the website \href{https://enzo-project.org}{enzo-project.org} for further details. For the simulation suite presented in this article, the public repository \href{https://github.com/SimonCSelg/enzo-dev\_mhdsgs-diskgalaxy}{github.com/SimonCSelg/enzo-dev\_mhdsgs-diskgalaxy}, commit \texttt{fc68caa}, was used. In order to maintain consistency between the simulations, this repository was frozen. Subsequent modifications are publicly available at \href{https://github.com/wolfram-schmidt/enzo-dev/tree/diskgalaxy}{github.com/wolfram-schmidt/enzo-dev/tree/diskgalaxy}.} 
The code is MPI-parallelized, features N-body dynamics based on a second-order drift-kick-drift algorithm in combination with cloud-in-cell interpolation to compute the joint gravitational potential of gas and particles, and a variety of different finite volume solvers for gas dynamics and magnetic fields. In our simulations, we applied the monotonic upstream-centered scheme for conservation laws with a local Lax-Friedrichs solver for sufficient numerical robustness. Adiabatic gas dynamics is applied under the assumption that magnetic field dynamics during interactions is mainly driven by tidal forces. Since radiative cooling, star formation, and feedback processes are not included, the spatial resolution and time stepping requirements are significantly reduced. Even so, the required computational resources for an extensive parameter study were substantial, as detailed below.

Originally, the computation of initial conditions for a disk in hydrostatic equilibrium were part of the initialization in Enzo \citep{Rodenbeck2016}. However, this was not practical for the particle initial conditions required for the live dark-matter halo. For this reason, we implemented a Python tool for computing particle initial conditions along the lines of ICICLE \citep{Drakos17}. By applying object-oriented programming, we were readily able to incorporate equilibrium disks as a subclass of the halo class into our Python code. The algorithms outlined in Sect.~\ref{sec:models} are implemented as object methods and make use of the highly optimized numerical methods for root finding and integration from the SciPy library. The resulting disk and particle data can be saved to files, which in turn can be read by Enzo to set up one or more disk galaxies embedded in live dark-matter halos.\footnote{The Python modules for halos and equilibrium disks are publicly available at \href{https://github.com/wolfram-schmidt/isolated-galaxy}{github.com/wolfram-schmidt/isolated-galaxy}.}

We restricted ourselves to interactions of Milky Way-like galaxies of equal mass. The parameters chosen for our simulations are summarized in Table~\ref{tab:model}. Each galaxy initially consists of a gas disk of $10^{10}\;\mathrm{M}_{\odot}$  and a dark-matter halo of $M_{200}=10^{12}\;M_{\odot}$, where $M_{200}$ is the mass enclosed within the virial radius $r_{200}=210\;$kpc. The radial scale length of the disks is $r_\mathrm{d}=3.5\;$kpc. The intial conditions were computed using the models outlined in Section~\ref{sec:models}. The initial separation $d_{\mathrm{sep},0}$ of the two galaxies is set to the cut-off radius $r_{200}$ of a dark-matter halo. For a periodic box of $2\;$Mpc linear size, this choice ensures that the separation of the centers of mass is small compared to the distance from the periodic boundaries. Although the halos are overlapping, the initial separation of the gas disks is significantly larger than their size. For our production runs, we used a $512^3$ root grid and 8 levels of refinement by overdensity and shear, corresponding to a maximal spatial resolution of $15\;$pc. The application of the second refinement criterion is important to cover interaction zones with high resolution, especially between the galaxies and in the outer disk regions where the gas density is relatively low. However, this greatly increases the computational cost of the simulations.\footnote{Using about 1200 MPI tasks, typically $1\;$million core-h were needed to complete one simulation. Owing to inherent limitations of the parallelization scheme used in Enzo, it is not feasible to run the code with more than a few thousand tasks.} 

The parameter space of galaxy interactions is spanned by
\begin{itemize}
\item the initial translational velocities of the two disks, $\vec{V}_1$ and $\vec{V}_2$ ,
\item the impact parameter, $b$, which is the normal distance between a straight line in the direction of the initial velocity of the secondary galaxy, $\vec{V}_2$, and the center of the primary galaxy,
\item and the disk orientations defined by the angular momentum vectors $\vec{L}_1$ and $\vec{L}_2$.
\end{itemize}
In realistic interaction scenarios, galaxies move on trajectories of various shapes and orientations within clusters and filaments. As long as we consider a system of two interacting galaxies in analogy to the two-body problem and ignore their environment, it is only the motion of the primary relative to the secondary that matters \citep{Toomre72,Moreno15}. The relative velocity is increasingly dominated by the gravitational potential as the galaxies approach each other. Consequently, the magnitudes of the initial velocities $\vec{V}_{1,2}$ of the galaxies are not overly important and can be estimated from the halo mass and initial separation.

In our simulation suite, we examined a series of impact parameters combined with varying orientations of the galaxies. To specify the orientations of the galaxies, we used the inclination angles of each galaxy's angular momentum vector with respect to the vector of the initial relative velocity of the interacting system, $\vec{V}_{\mathrm{rel}} = \vec{V}_{1} - \vec{V}_{2}$:
\begin{equation}
    i_{1,2} = \angle(\vec{L}_{1,2}, \vec{V}_{\mathrm{rel}}),
\end{equation}
The relative inclination of the disks is given by the difference $|i_{2} - i_{1}|$.

\begin{table*}
\caption{Overview of varied simulation parameters and minimum distances between galaxy centers.}
\label{tb:parameters}
\centering
\begin{tabular}{ccrrrrrr}
  \hline\hline
  \multirow{2}{*}{Group}
  & \multirow{2}{*}{\simno} 
    & \multicolumn{3}{c}{Parameters $[^{\circ}]$} 
    & \multicolumn{3}{c}{Minimum distance $[\mathrm{kpc}]$} \\  \cline{3-8}
  & & $i_1$ & $i_2$ & $\alpha_{\mathrm{b}}$ & $1^{\mathrm{st}}\;$encounter & $2^{\mathrm{nd}}\;$encounter & $3^{\mathrm{rd}}\;$encounter \\  
  \hline
  \multirow{6}{*}{0}
  & 1 & 90 & 90 & 0 & 0.00 & 0.00 & 6.44 \\
  & 2 & 90 & 90 & 15 & 7.41 & 0.00 & 0.41 \\
  & 3 & 90 & 90 & 20 & 10.88 & 2.06 & 0.18 \\
  & 4 & 90 & 90 & 25 & 14.27 & 1.79 & 2.32 \\
  & 5 & 90 & 90 & 30 & 18.80 & 2.67 & 3.04 \\
  & 6 & 90 & 90 & 45 & 31.33 & 7.74 & 0.00 \\
  \hline
  \multirow{4}{*}{I}
  & 7 & 90 & 135 & 0 & 1.69 & 1.16 & 2.23 \\
  & 8 & 90 & 135 & 15 & 7.15 & 0.00 & 2.29 \\
  & 9 & 90 & 135 & 30 & 18.15 & 2.83 & -- \\
  & 10 & 90 & 135 & 45 & 30.95 & -- & -- \\
  \hline
  \multirow{4}{*}{II}
  & 11 & 90 & 0 & 0 & 1.40 & 3.17 & 0.00 \\
  & 12 & 90 & 0 & 15 & 7.33 & 0.00 & --   \\
  & 13 & 90 & 0 & 30 & 18.32 & 0.00 & -- \\
  & 14 & 90 & 0 & 45 & 31.29 & -- & -- \\
  \hline
  \multirow{4}{*}{III}
  & 15 & 0 & 0 & 0 & 0.57 & 1.62 & 0.82 \\
  & 16 & 0 & 0 & 15 & 7.37 & 0.86 & 0.18 \\
  & 17 & 0 & 0 & 30 & 18.42 & 2.83 & -- \\
  & 18 & 0 & 0 & 45 & 31.11 & 7.64 & 2.44 \\
  \hline
  \multirow{4}{*}{IV}
  & 19 & 67.5 & 112.5 & 0 & 0.10 & 0.00 & 1.32 \\
  & 20 & 67.5 & 112.5 & 15 & 6.26 & 0.00 & 0.84 \\
  & 21 & 67.5 & 112.5 & 30 & 14.41 & 6.68 & -- \\
  & 22 & 67.5 & 112.5 & 45 & 24.12 & -- & -- \\
  \hline
  \multirow{4}{*}{V}
  & 23 & 45 & 135 & 0 & 4.55 & 5.49 & 1.12 \\
  & 24 & 45 & 135 & 15 & 6.86 & 0.00 & 0.50 \\
  & 25 & 45 & 135 & 30 & 7.36 & 2.14 & 2.00 \\
  & 26 & 45 & 135 & 45 & 12.43 & 0.00 & -- \\ 
  \hline
\end{tabular}
\tablefoot{The parameters are the inclination angles $i_1$ and $i_2$ of the disk's rotation axes measured with respect to their initial relative velocity $\vec{V}_{\mathrm{rel}}$, and the angle $\alpha_{\mathrm{b}}$ defined by $\sin{\alpha_{\mathrm{b}}} = b/d_{\mathrm{sep},0}$, where $b$ is the impact parameter and $d_{\mathrm{sep},0}$ the initial separation of the two disks. Dashes correspond to missing data in the case of simulations that were not evolved long enough to undergo more than one or two encounters.}
\end{table*}

In Table~\ref{tb:parameters}, we provide an overview of the parameters of all production runs. For referencing, the simulations are numbered sequentially (second column). In addition, simulations are grouped according to the disk inclinations (first column):
\begin{enumerate}
    \item Group 0: edge-on collisions with parallel rotation axes perpendicular to relative motion\\ ($i_{1}=i_{2}=90^{\circ}$).
    \item Group I: same orientation of primary, secondary inclined by $i_{2}=135^{\circ}$ ($|i_{2} - i_{1}| = 45^{\circ}$).
    \item Group II: same orientation of primary, secondary's rotation axis aligned with relative motion ($i_{1}=90^{\circ}$, $i_{2}=0^{\circ}$, $|i_{2}-i_{1}| = 90^{\circ}$).
    \item Group III: face-on collisions with relative velocity parallel or antiparallel to rotation axes\\ ($i_{1}=i_{2}=0^{\circ}$).
    \item Group IV: primary and secondary inclined by $i_{1}=67.5^{\circ}$ and $i_{2}=112.5^{\circ}$, respectively\\ ($|i_{2}-i_{1}| = 45^{\circ}$).
    \item Group V: primary and secondary inclined by $i_{1}=45^{\circ}$ and $i_{2}=135^{\circ}$, respectively\\ ($|i_{2} - i_{1}|=90^{\circ}$).
\end{enumerate}
In each group, the impact parameter $b$ is varied between the limiting scenarios of a central collision ($b=0$) and an interaction where the galaxies and their halos are moderately distorted by tidal forces ($b$ on the order of virial radius). In the following, we use the angle defined by $\sin{\alpha_{\mathrm{b}}} = b/d_{\mathrm{sep,0}}$ to specify the impact parameter. For the absolute value of the initial relative velocity, we assumed $|\vec{V}_{\mathrm{rel}}|= v_{200}$ in all cases. 

Owing to the required computational time and substantial storage requirements, only a few dozen scenarios with more or less arbitrary parameter choices were feasible. Thus, our simulation suite covers some typical scenarios, such as edge-on (group 0) and face-on (group III) collisions of aligned disks, and a selection of intermediate cases with nonzero relative inclinations of the disks. In groups IV and V, we investigated scenarios where the secondary galaxy moves along an inbound trajectory that is inclined with respect to the disk plane of the primary (similar to satellites of the Milky Way, such as the Magellanic Clouds). If feasible, the simulations were run until the post-merger stage. In group 0, this was achieved for \simno~1 to~5 within $2.5\;$Gyr, but not for \simno~6. In the other groups, reductions of the
fraction of the time step implied by the Courant-Friedrichs-Lewy criterion required terminating the simulations already after $1.5\;$Gyr in order to save computational time. However, all simulations captured at least the first interaction between the galaxies. 


\begin{figure*}
   \centering
   \includegraphics[width=0.48\textwidth]{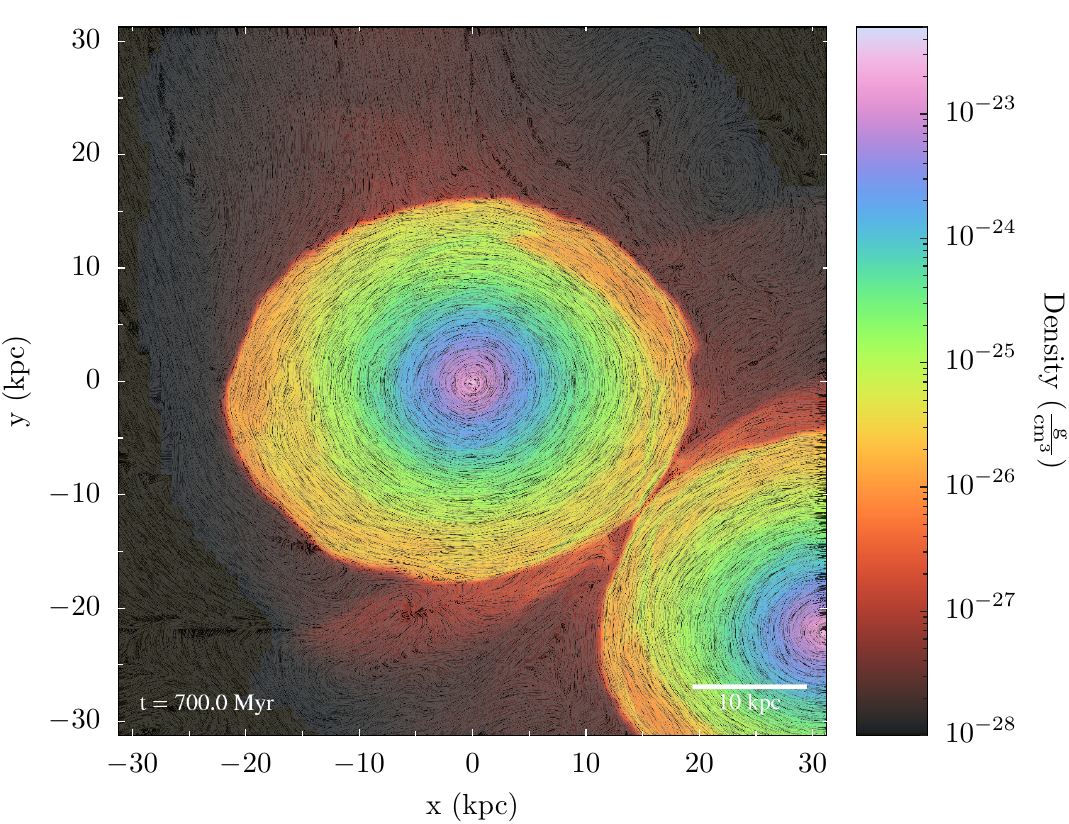}
   \includegraphics[width=0.48\textwidth]{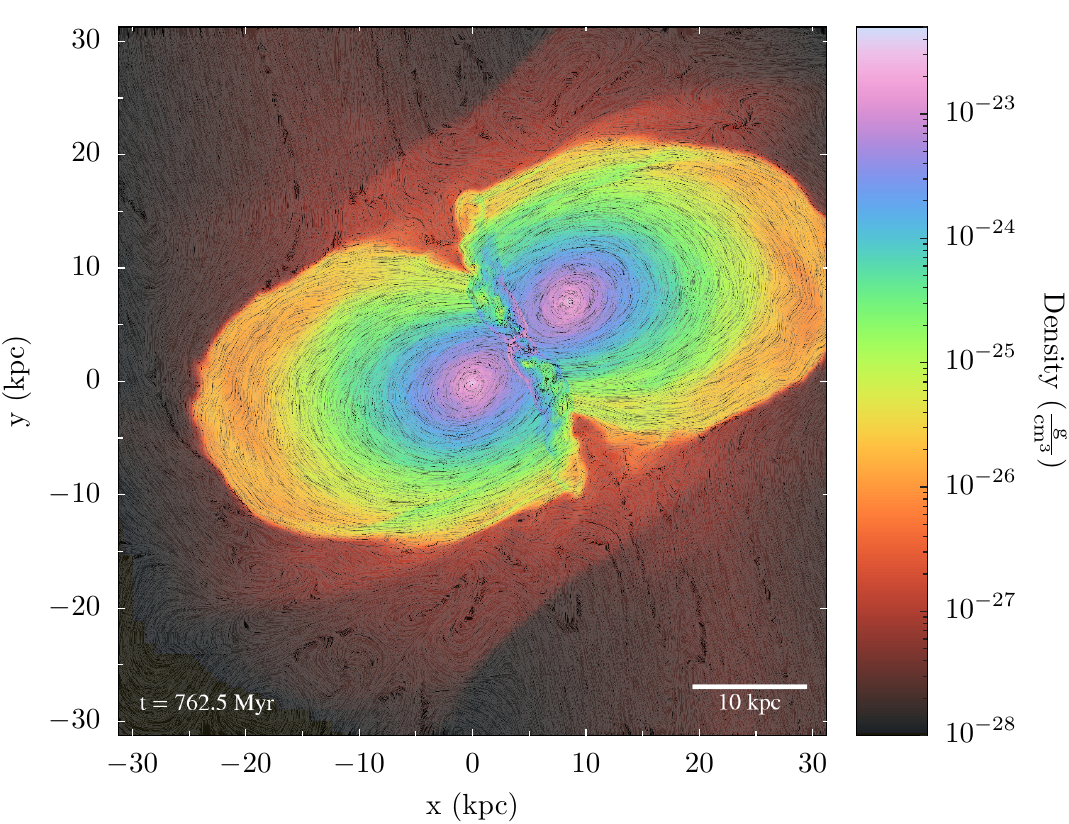}
   \includegraphics[width=0.48\textwidth]{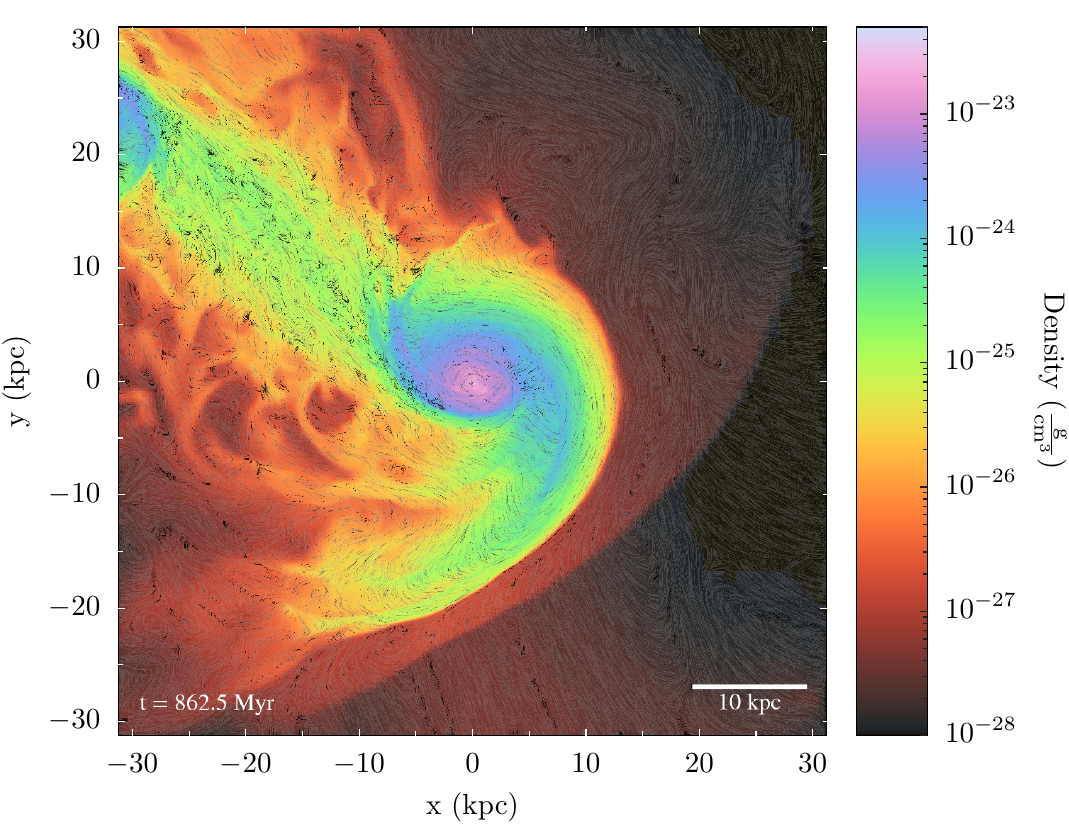}
   \includegraphics[width=0.48\textwidth]{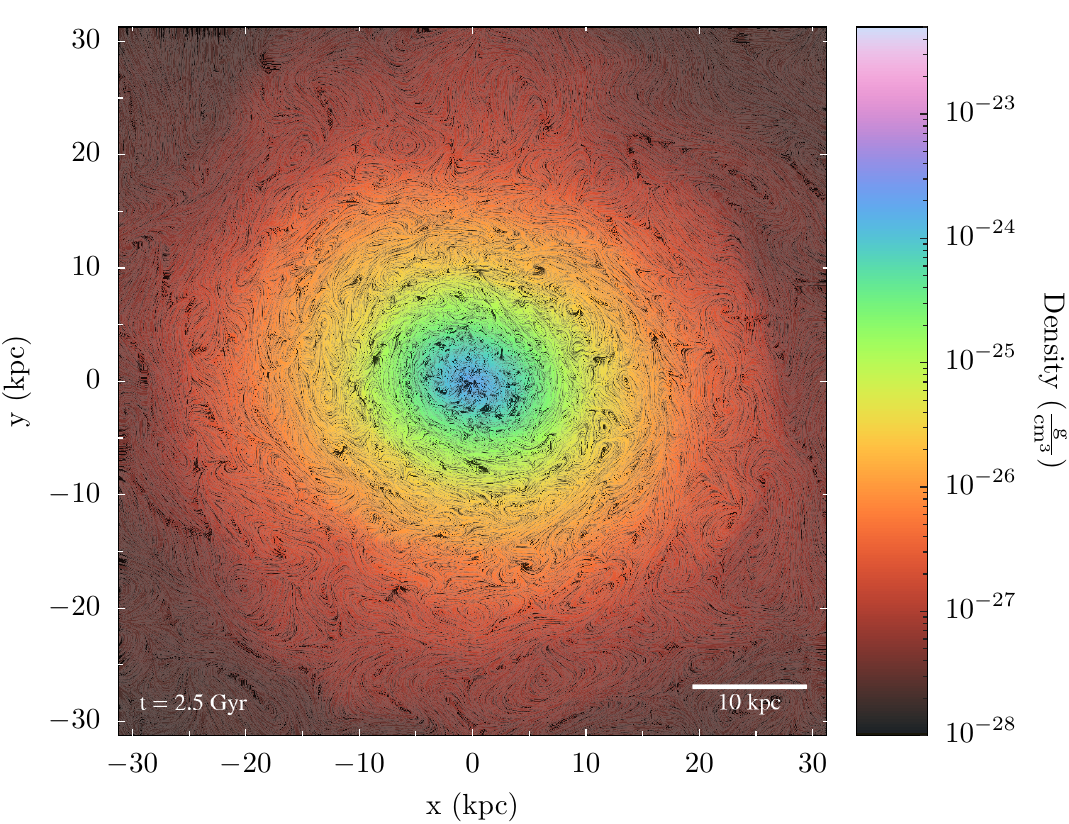}
   \caption{Slices of the gas density in combination with a line integral convolution of the magnetic field in an edge-on collision of two disk galaxies (\simno~3 in Table~\ref{tb:parameters}). The first three panels show stages around the first encounter ($t=762.5\;$Myr). At the end of the simulation ($t=2.5\,$Gyr), the two galaxies have merged. }
   \label{fig:collision}
\end{figure*}

\begin{figure}
    \centering
    \includegraphics[width=\linewidth]{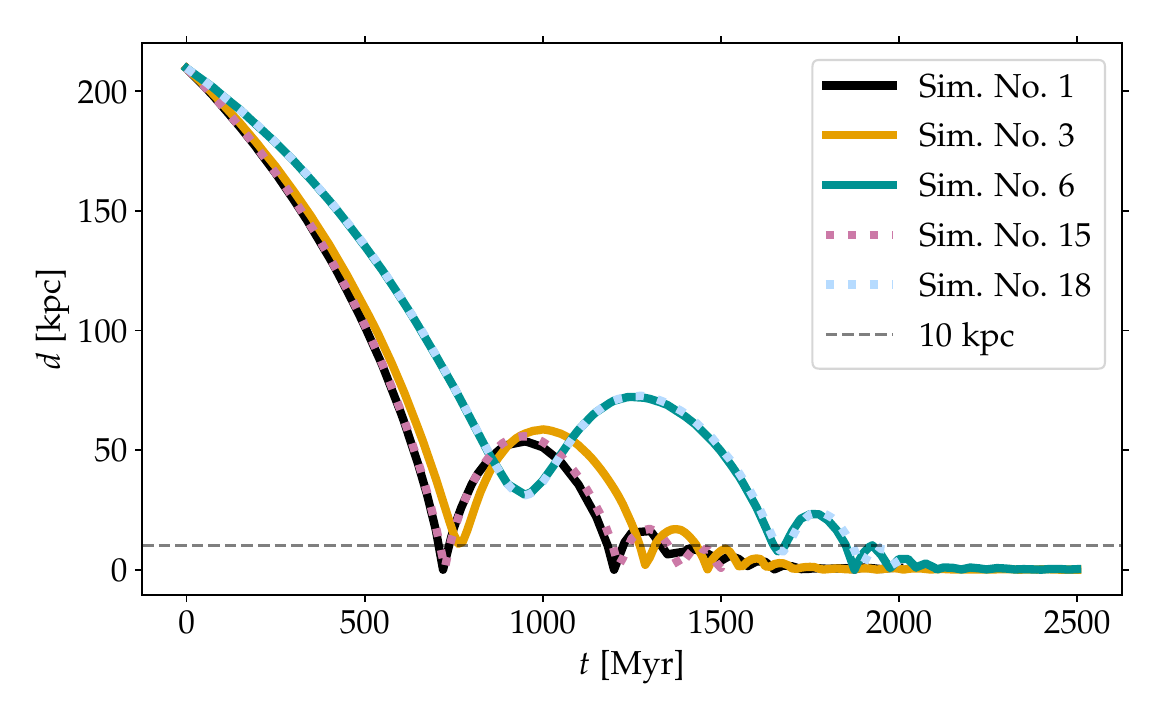}
    \caption{Time evolution of the separation, $d$, between the centers of interacting galaxies for five cases. \simno 1 and 15 are central collisions ($\alpha_{\mathrm{b}}=0^{\circ}$) at different initial inclinations (see Table~\ref{tb:parameters}). For \simno~3 ($\alpha_{\mathrm{b}}=20^{\circ}$), different interaction stages are visualized in Fig.~\ref{fig:collision}. \simno~6 and~18 are examples for grazing interactions at $\alpha_{\mathrm{b}}=45^{\circ}$. The dashed gray line marks a separation of $10\;$ kpc, corresponding to the diameter of the center regions of the initial disks.}
    \label{fig:separation}
\end{figure}

\begin{figure}
    \centering
    \includegraphics[width=0.48\textwidth]{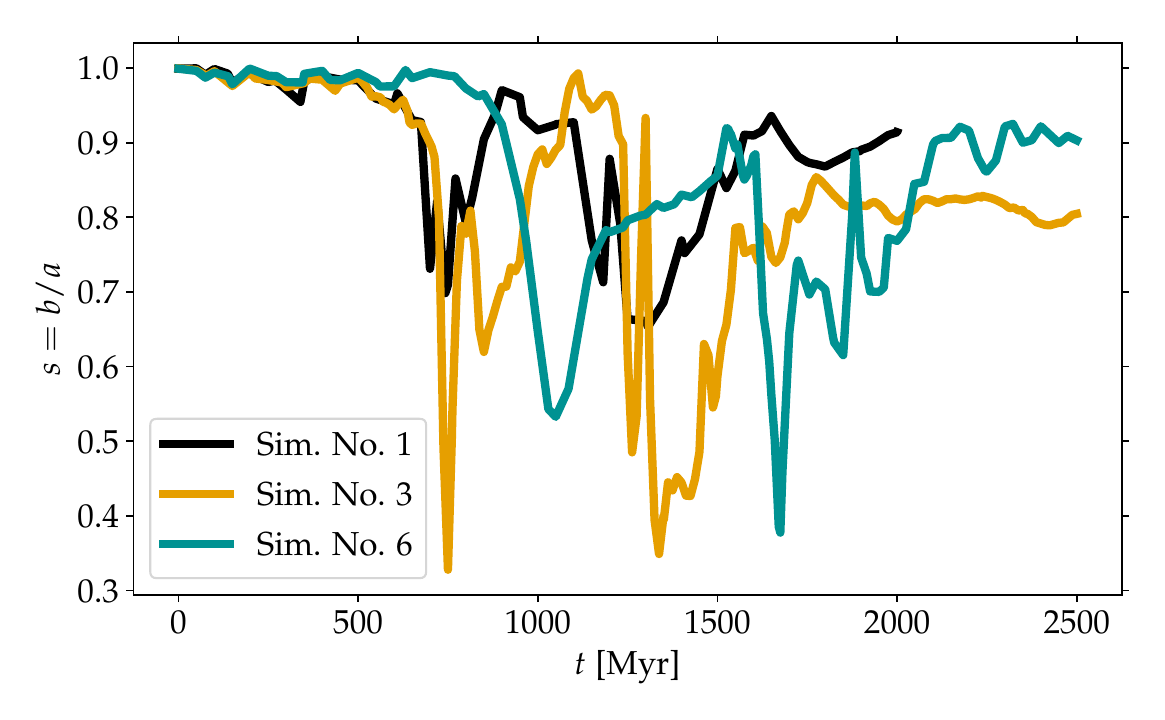}
    \includegraphics[width=0.48\textwidth]{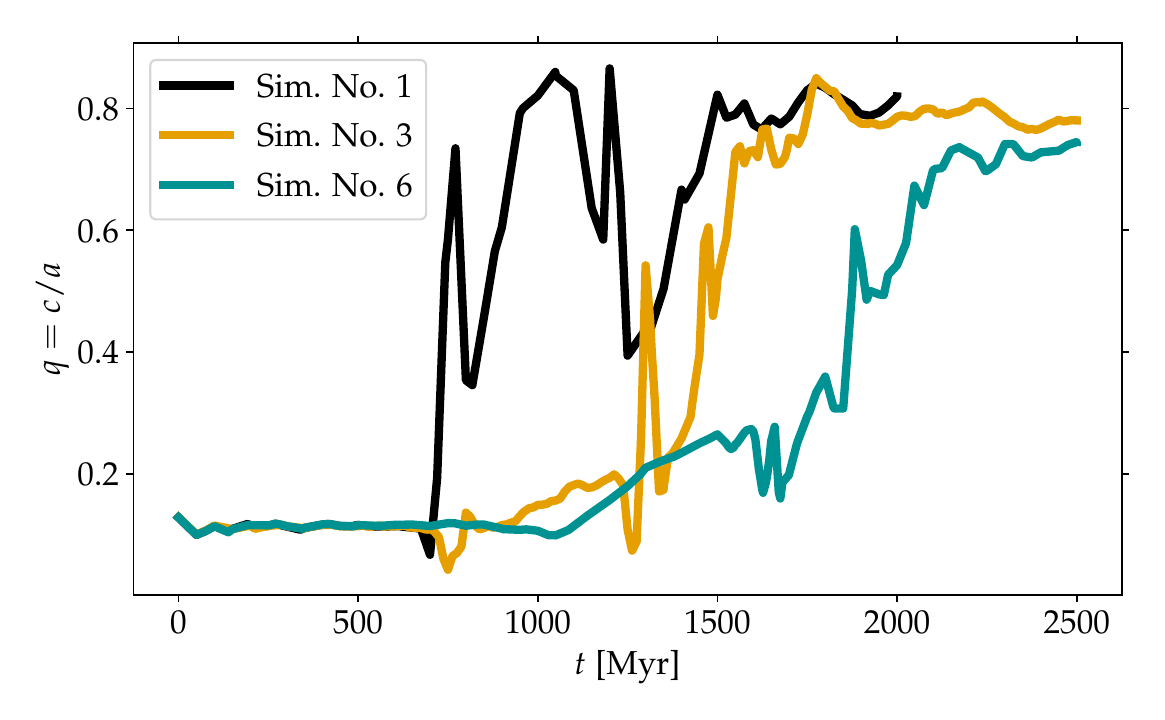}
    \caption{Evolution of galaxy shape parameters $s = b/a$ (\emph{left}) and $q = c/a$ (\emph{right}) of \simno~1, 3, and 6, which represent a sequence of simulations with increasing impact parameter for fixed initial inclinations. }
    \label{fig:ShapeFixedInclination}
\end{figure}

\section{Results}
\label{sec:results}

\subsection{Phenomenology of interacting disks}
\label{sec:phenomenology}

We begin with a phenomenological discussion of representative examples. In Fig.~\ref{fig:collision}, density slices from an edge-on collision of two galaxies with $\alpha_{\mathrm{b}} = 20^{\circ}$ (\simno~3 in Table~\ref{tb:parameters}) are overlaid with a visualization of magnetic field lines using the line integral convolution method.\footnote{Like iron filings surrounding a bar magnet, magnetic field lines are visualized by filtering a texture consisting of white-noise along local streamlines derived from the magnetic field data.} The first three panels in Fig.~\ref{fig:collision} show the disks shortly before the collision, the time of closest approach (pericentric passage), and $100\,$Myr later. The separation of the galaxy centers is plotted as a function of time in Fig.~\ref{fig:separation}. After reversing their motion, the galaxies are pulled toward each other once more and finally merge into a remnant of elliptical shape (Fig.~\ref{fig:collision}, lower right). This is indicated by the time evolution of the shape parameters $s$ and $q$ plotted in Fig.~\ref{fig:ShapeFixedInclination}. The two parameters specify the ratios of the three principal axes (thin disk: $s=1$, $q\sim 0$; sphere: $q=s=1$). As can be seen, the initially disk-like galaxies are strongly distorted during the interaction stages and evolve into an ellipsoid in the post-merger phase.

The gas density indicated by the color map in Fig.~\ref{fig:collision} shows that a substantial amount of gas is ripped off the initial disks by tidal forces. The magnetic field structure reveals that the initially ordered toroidal fields inside the two disks evolves into highly turbulent structures expanding into the surroundings of the galaxies. After the first encounter, this can be mainly observed in the region in between the two disks (the so-called bridge) and in the tidal tails. The final remnant exhibits an unordered field that extends far into low-density regions. This suggests that interactions play a role in the magnetization of the circumgalactic medium, although the magnetic field in a realistic cosmological environment is considerably larger than the weak background field of $10^{-25}\;$G assumed in our simulations.

\begin{figure*}
\centering
\includegraphics[width=0.49\textwidth]{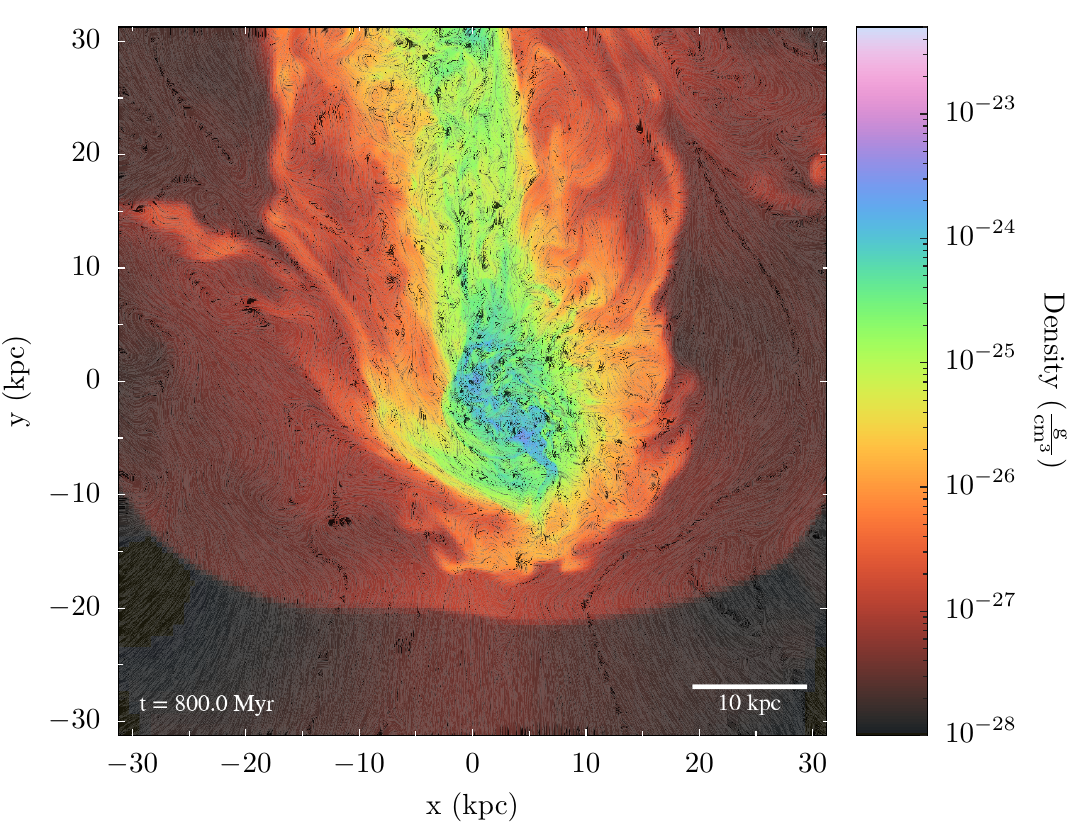}
\includegraphics[width=0.49\textwidth]{figures/0B20_data0048_density_slice_z_32_high_center-windowmagnetic_f-scaled_by_1000e-31_5000e-26-file_based_tracked_tight-layout_kamae_0_magnetic_field.pdf}
\includegraphics[width=0.49\textwidth]{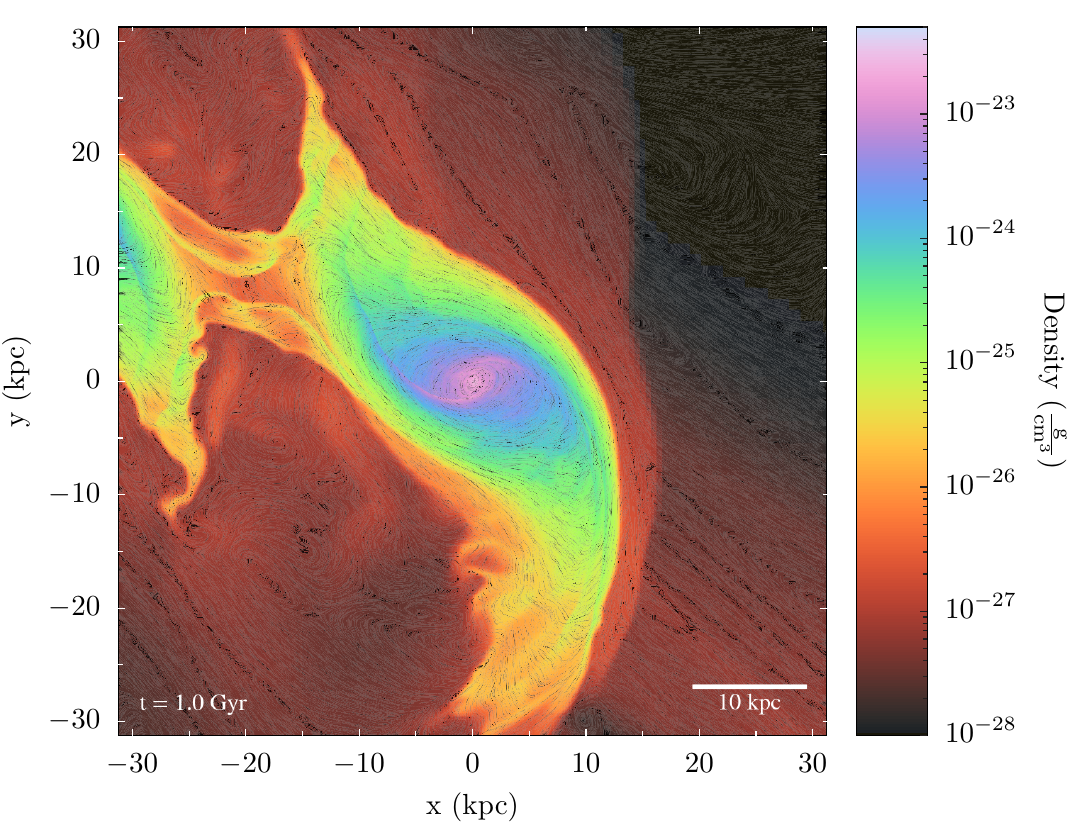}
\includegraphics[width=0.49\textwidth]{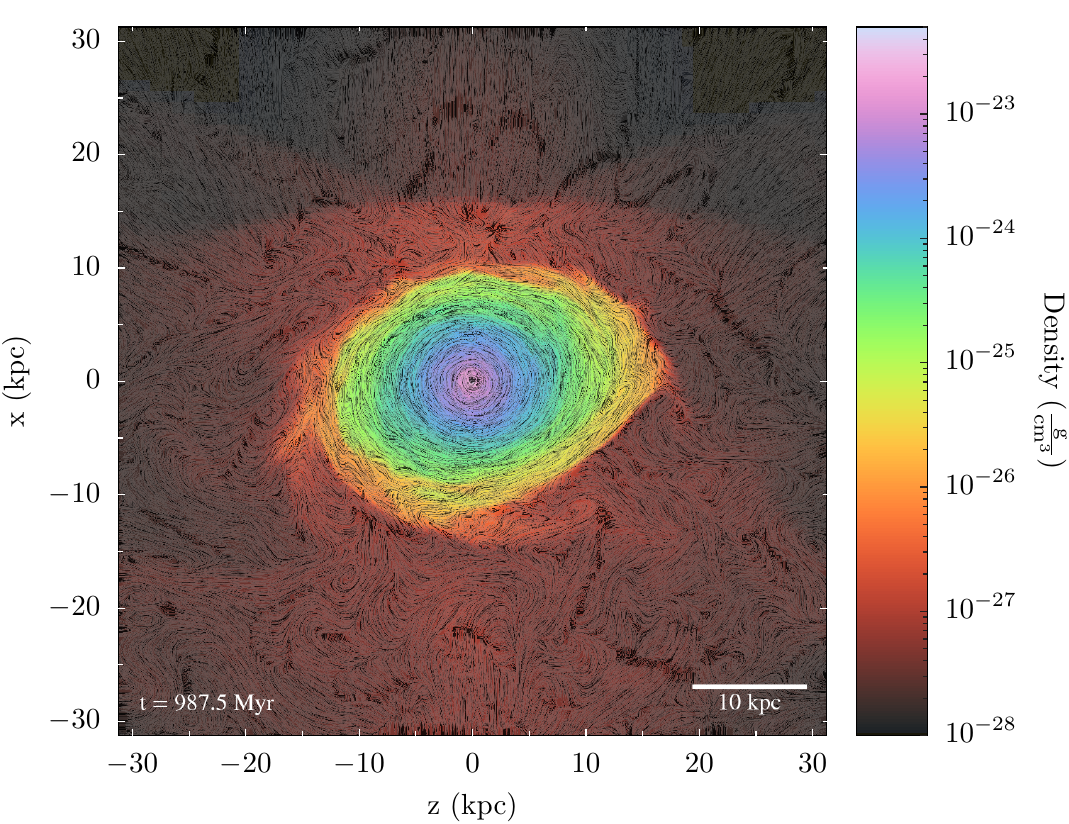}
\caption{Slices of the gas density in combination with a line integral convolution of the magnetic field showing post-interaction stages for four different initial orientations. From left to right and top to bottom: \simno~1, 3, 6, and 18 (see Table~\ref{tb:parameters}). While the disk inclinations are the same in the first three simulations, \simno~6 and 18 use the impact parameter $\alpha_\mathrm{b}=45^\circ$.}
\label{fig:LicPostInteraction}
\end{figure*}

For nearly central collisions with a minimal separation below $10\;\mathrm{kpc}$, the disks are strongly perturbed after the first interaction (Fig.~\ref{fig:LicPostInteraction}, upper panels). With increasing impact parameter, only the outer regions of the disks interact and less turbulence is produced. A bridge connecting the disks and tidal tails become the most prominent interaction features in the case of a grazing encounter (Fig.~\ref{fig:LicPostInteraction}, lower left). Although the magnetic field remains mostly ordered at this stage, the disks eventually undergo substantial changes and finally merge, as indicated by the time evolution of the separation and shape parameters plotted in Figs.~\ref{fig:collision} and~\ref{fig:ShapeFixedInclination}, respectively. The distortion of the disks is reduced further if angular momentum is aligned with velocity ($i_1=i_2=0^{\circ}$), that is to say, if\ the disk planes of both galaxies are oriented perpendicular to the velocity vector $\vec{V}_{\mathrm{rel}}$ (Fig.~\ref{fig:LicPostInteraction}, lower right). This type of collision is called face-on (group III in Table~\ref{tb:parameters}). Consequently, not only the impact parameter, but also the disk orientation has a significant influence on the outcome of galaxy interactions.

\begin{figure*}
    \centering
    \includegraphics[width=0.49\textwidth]{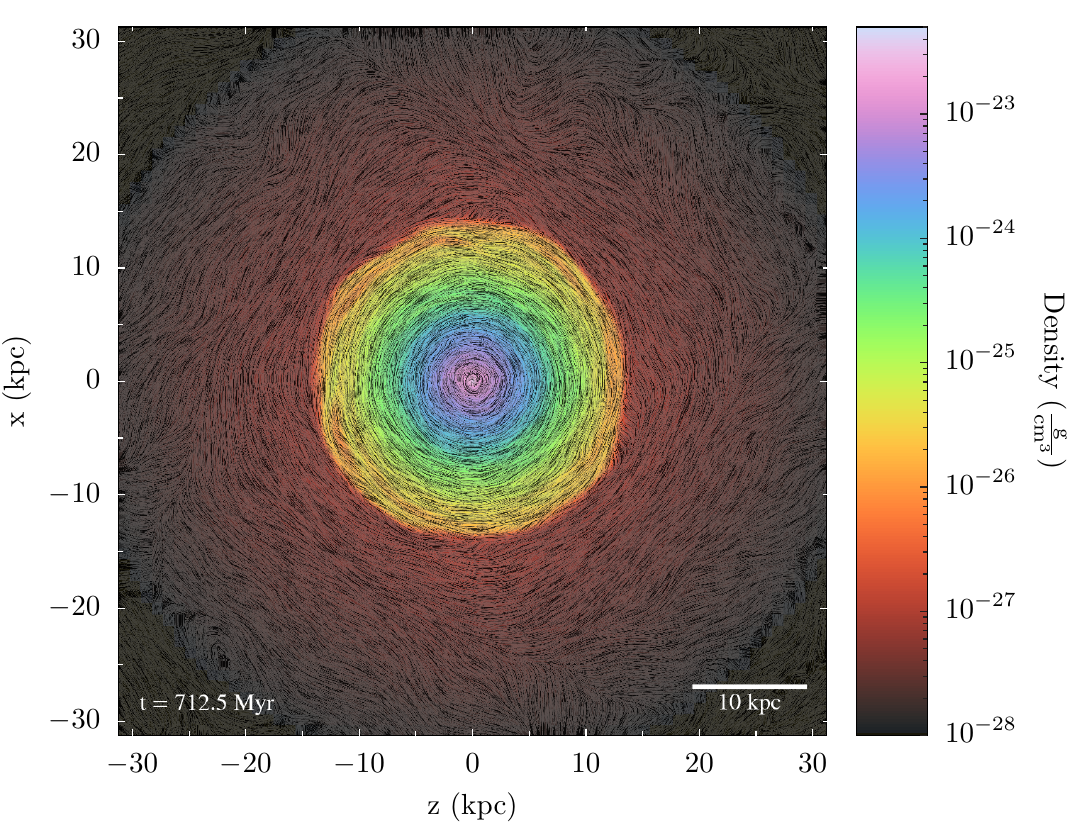}
    \includegraphics[width=0.49\textwidth]{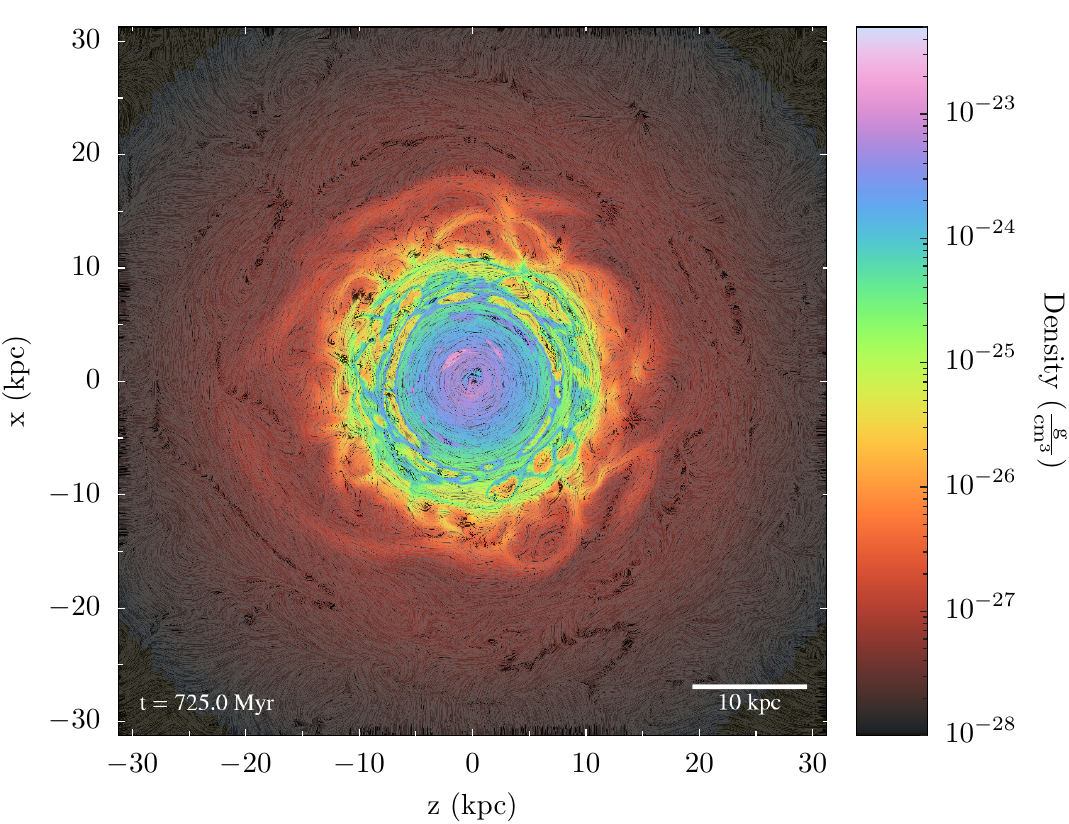}
    \includegraphics[width=0.49\textwidth]{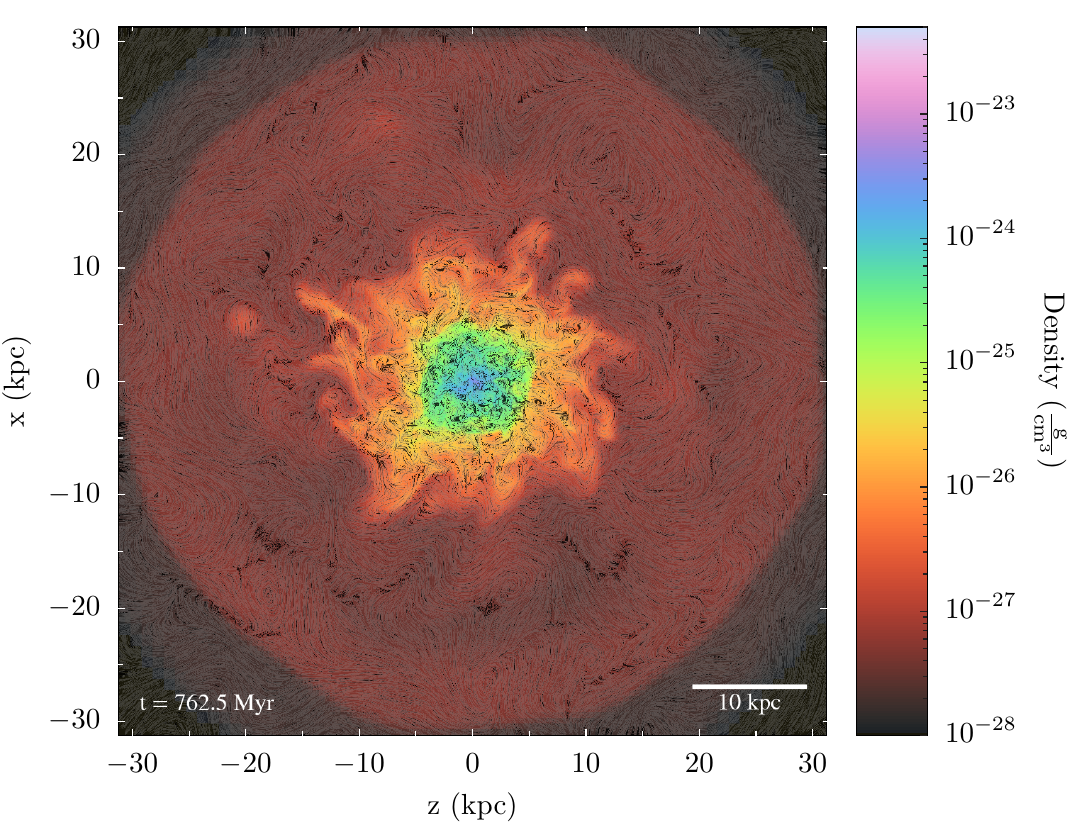}
    \includegraphics[width=0.49\textwidth]{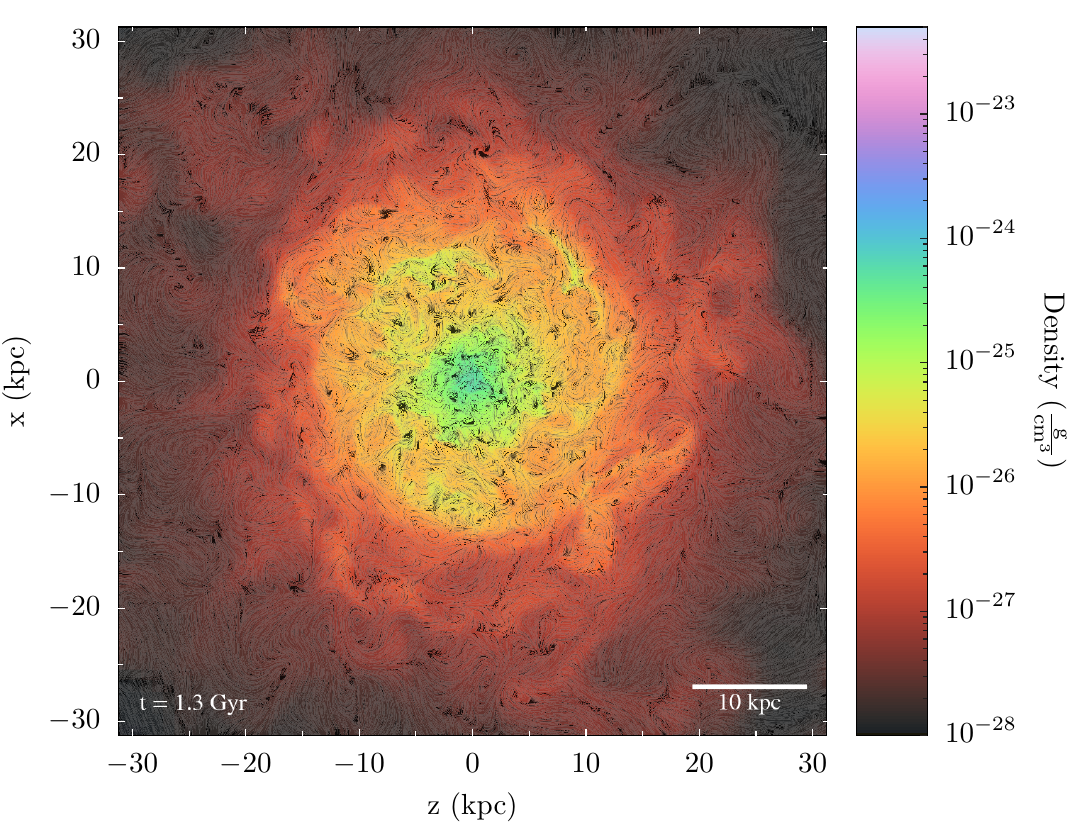}
    \caption{Similar stages as in Fig.~\ref{fig:collision} for a simulation with impact parameter $\alpha_{\mathrm{b}}=0^{\circ}$ and inclination angles $i_1=i_2=0^{\circ}$ (\simno~15). }
    \label{fig:FaceOnCollision}
\end{figure*}

Another example of a face-on collision from group III is shown in Fig.~\ref{fig:FaceOnCollision}. In this simulation, the impact parameter is zero (\simno~15). The galaxies have parallel angular momentum vectors (i.e., the rotation of the two disks is prograde). The morphology after the first encounter is markedly different from the central edge-on collision shown in Fig.~\ref{fig:LicPostInteraction} (upper left). Instead of tidal tails and a highly turbulent bridge, ring-like perturbations in the outer disks are accompanied by braid-like outward gas streams. As time progresses, more gas is scattered outward and most of the outer disk disperses (Fig.~\ref{fig:FaceOnCollision}, lower left). The remnant after about $1.5\;$Gyr is highly irregular and does not resemble a disk any longer (Fig.~\ref{fig:FaceOnCollision}, lower right). In the following sections, we investigate whether the morphological differences between different types of interactions are reflected in magnetic field statistics. 


\begin{figure*}
    \centering
    \includegraphics[width=0.49\textwidth]{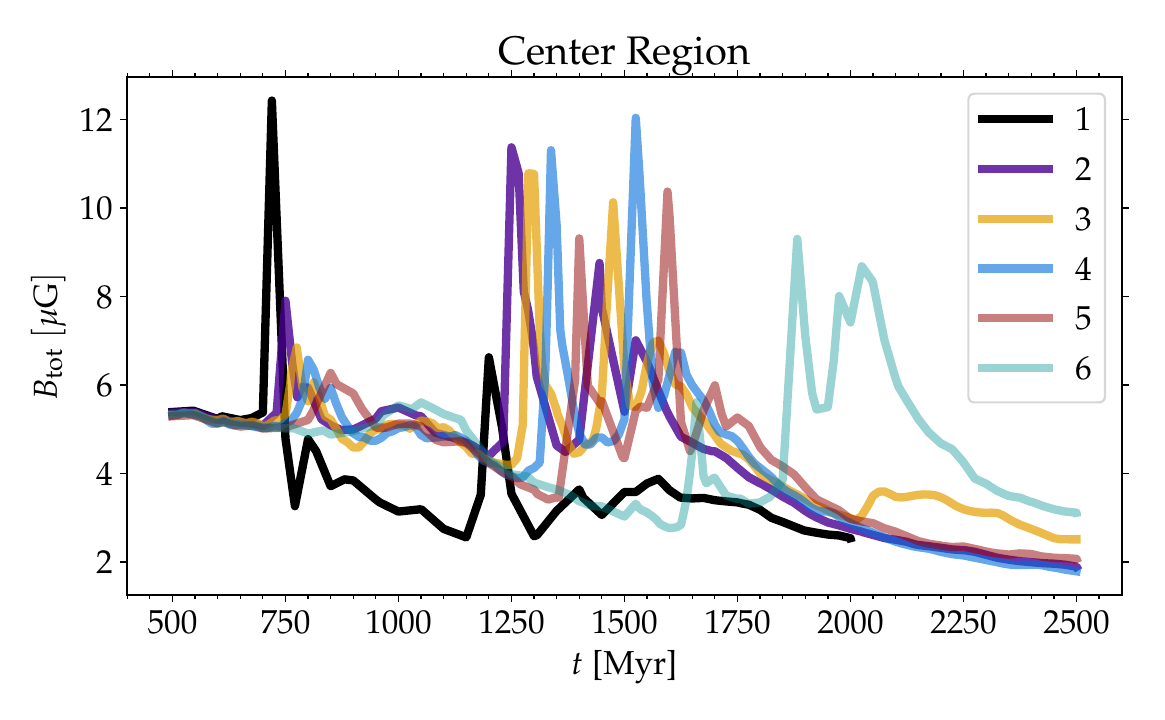}
    \includegraphics[width=0.49\textwidth]{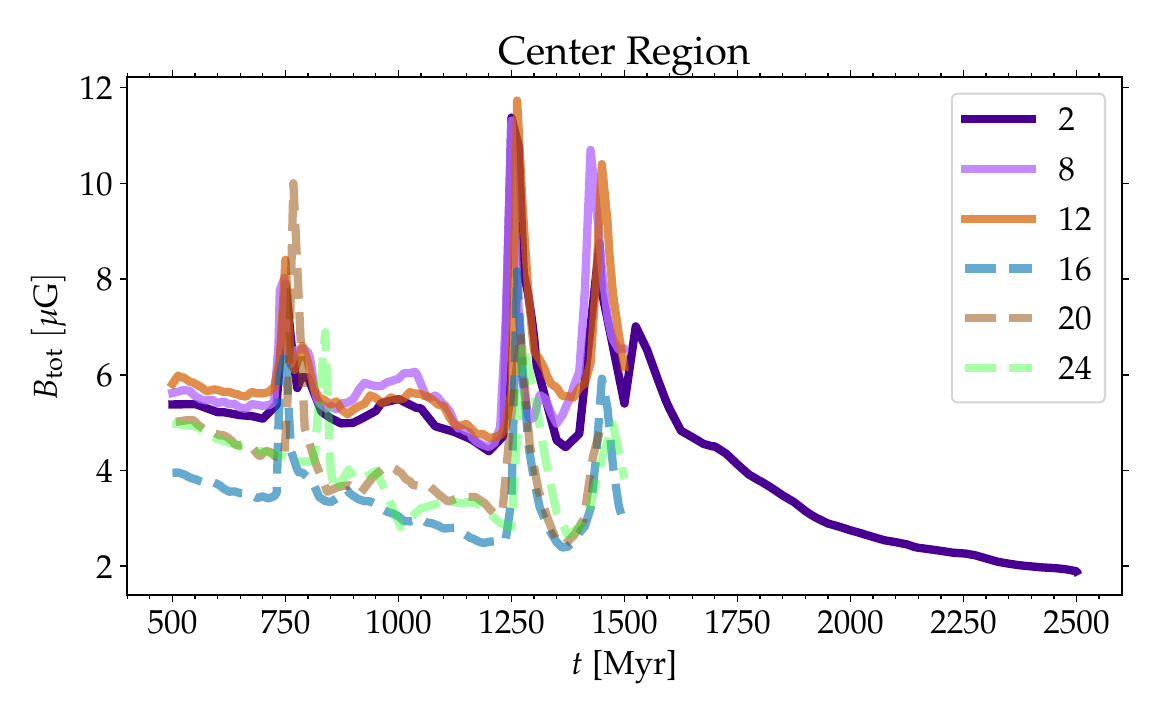} \\
    \includegraphics[width=0.49\textwidth]{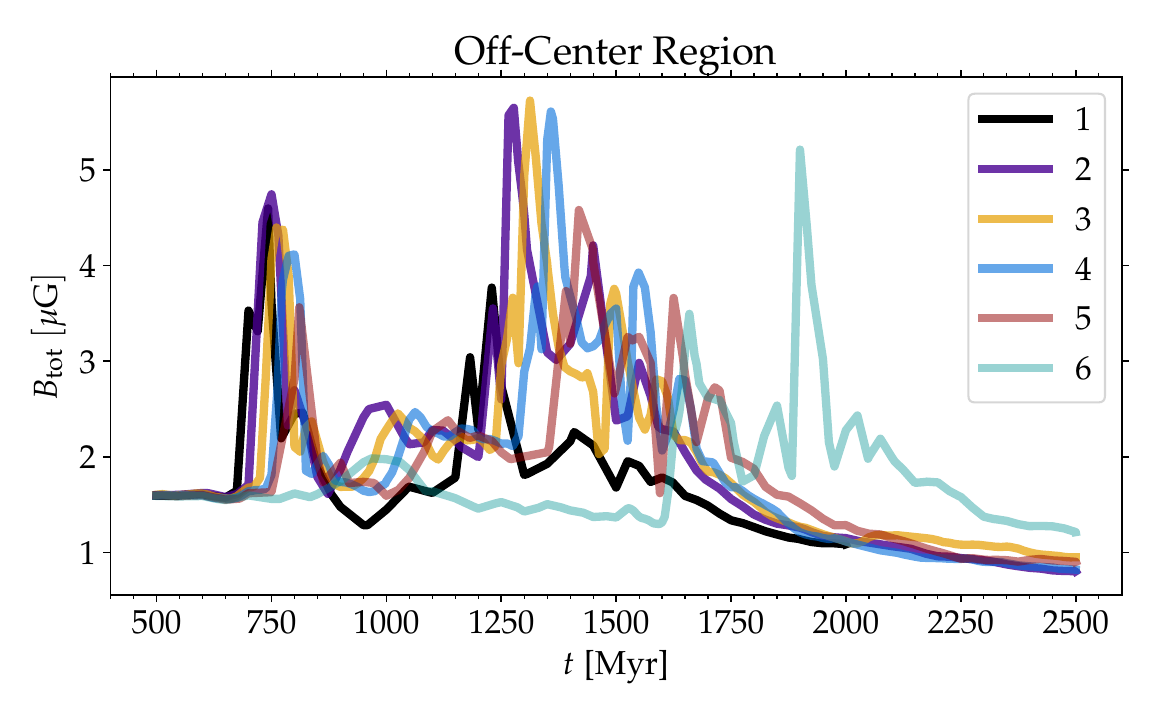}
    \includegraphics[width=0.49\textwidth]{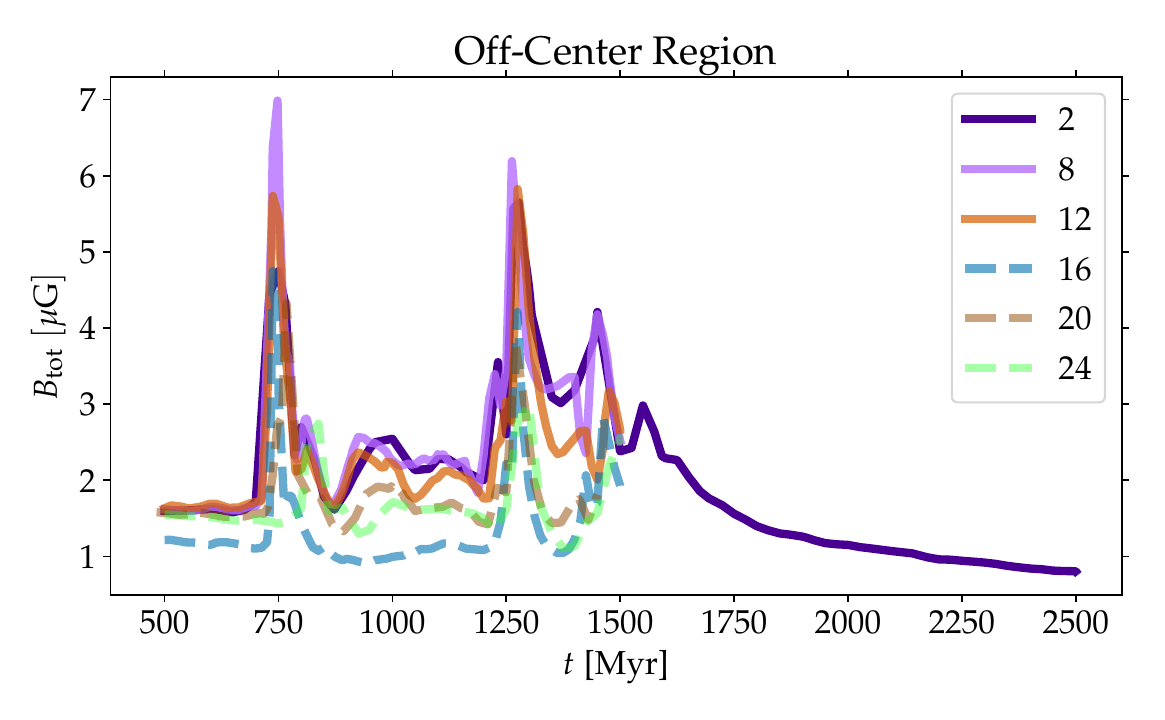}
    \caption{Evolution of the magnetic field strength averaged over center (\emph{top}) and off-center regions (\emph{bottom}). \emph{Left column}: Simulations of edge-on collisions ($i_1=i_2=90^{\circ}$). \emph{Right column}: Simulations with different disk inclinations sharing the impact parameter $\alpha_{\mathrm{b}}=15^{\circ}$. The simulation numbers indicated in the legends refer to Table~\ref{tb:parameters}.}
    \label{fig:FieldEvol}
\end{figure*}

\subsection{Average magnetic field evolution}

As a quantitative measure, we averaged the magnetic field strength in cylindrical regions enclosing the initial gaseous disks (see also \citealt{Rodenbeck2016}). These regions have a radius of $20\;$kpc and a height of $4\;$kpc. By additionally averaging over an inner disk of $5\;$kpc radial extent, we can distinguish between a central region ($r\le 5\;$kpc) and an off-center region ($5\;\mathrm{kpc} < r\le 20\;\mathrm{kpc}$).

Figure~\ref{fig:FieldEvol} shows the mean magnetic field strength in the primary galaxy as a function of time for different subsets of simulations. The amplification of the magnetic field by interactions is indicated by the height of peaks. Edge-on collisions (group 0) with impact parameter varying between $\alpha_\mathrm{b}=0^\circ$ and $45^\circ$ are shown in the left plots. While the collision is central for $\alpha_\mathrm{b}=0^\circ$, the impact parameter $b$ is about $150\;$kpc for $\alpha_\mathrm{b}=45^\circ$. As expected from the phenomenological discussion in Sect.~\ref{sec:phenomenology}, the first maximum of the average magnetic field in center regions shows a decline with increasing impact parameter (Fig.~\ref{fig:FieldEvol}, upper left). This indicates that the magnetic field in the inner disk is strongly amplified only for nearly central collisions. In contrast, differences between the average magnetic field in off-center regions are less pronounced because the outer disks are tidally disrupted even for relatively large impact parameters (Fig.~\ref{fig:FieldEvol}, lower left). The largest field strength in an off-center region is observed for $\alpha_\mathrm{b}=15^\circ$ (\simno~2). This suggests that the mixture of shock fronts and turbulence is particularly efficient in driving the amplification of the field if the collision is slightly off-center. During the second encounter, particularly strong fields are produced for $\alpha_\mathrm{b}$ in the range from $15^\circ$ and $\alpha_\mathrm{b}=25^\circ$ (\simno~2 to 4). For $\alpha_\mathrm{b}\ge 25^\circ$, the maximum field strength in center regions is reached in the third peak. These trends show that in the case of gracing interactions several approaches are required to affect the magnetic field throughout the disks.

Comparing simulations from different groups at a fixed impact parameter ($\alpha_\mathrm{b}=15^\circ$), we see that peaks approximately coincide, but the peak height varies with the disk orientation (right plots in Fig.~\ref{fig:FieldEvol}). Peaks are particularly pronounced if the primary's motion is parallel to its disk plane ($i_1=90^\circ$, groups 0, I, and II). For the face-on collision (\simno~16 in group III), magnetic field amplification is significantly reduced both in center and off-center regions. The average magnetic field just prior to the first encounter is also smaller in this case. This is a consequence of stronger gas stripping if the disk moves face-on relative to the ambient medium, which removes magnetic energy from the cylindrical volume in which averages are computed. In groups IV and V, both disks are inclined with respect to their relative motion and their relative inclinations are $|i_2-i_1|=45^\circ$ and $90^\circ$, respectively. In these two cases, the peaks are comparable to the face-on collision, with the exception of the center region in \simno~20 during the first encounter. Overall, our findings suggest that the most important parameter is the disk inclination (i.e., the angle between the disk velocity and rotation axis); the relative inclination $|i_2-i_1|$ also plays a role, but it is subdominant.
\begin{figure*}
    \centering
    \includegraphics[width=0.45\textwidth]{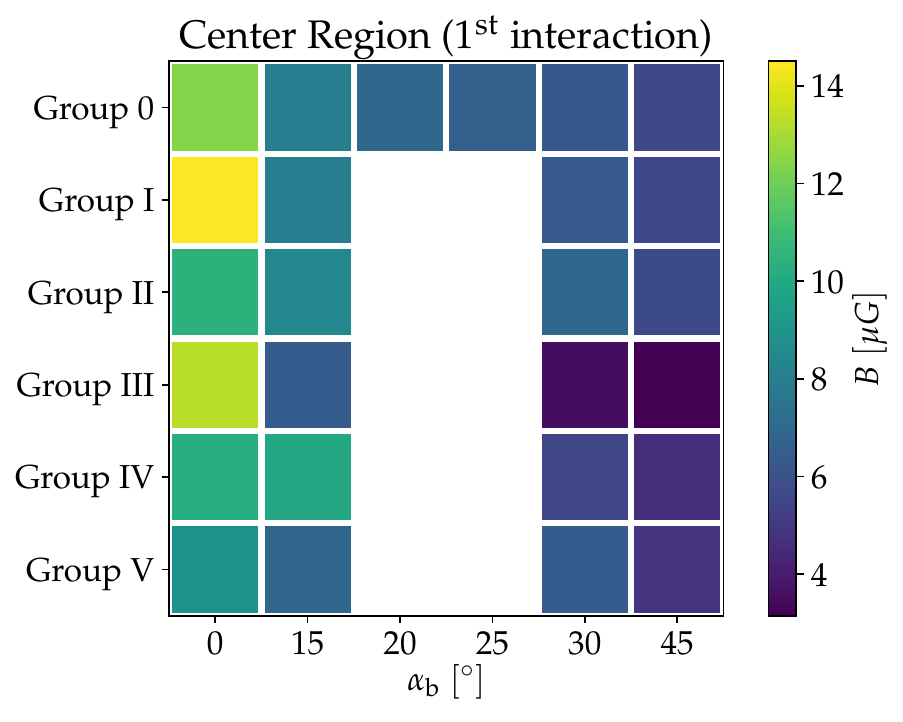}
    \includegraphics[width=0.45\textwidth]{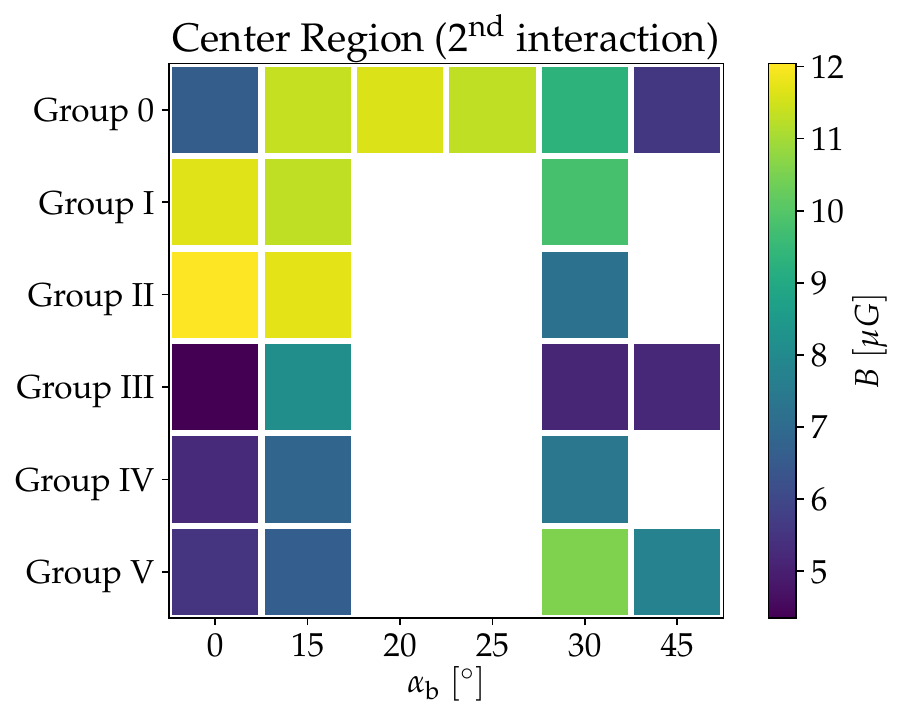} \\
    \includegraphics[width=0.45\textwidth]{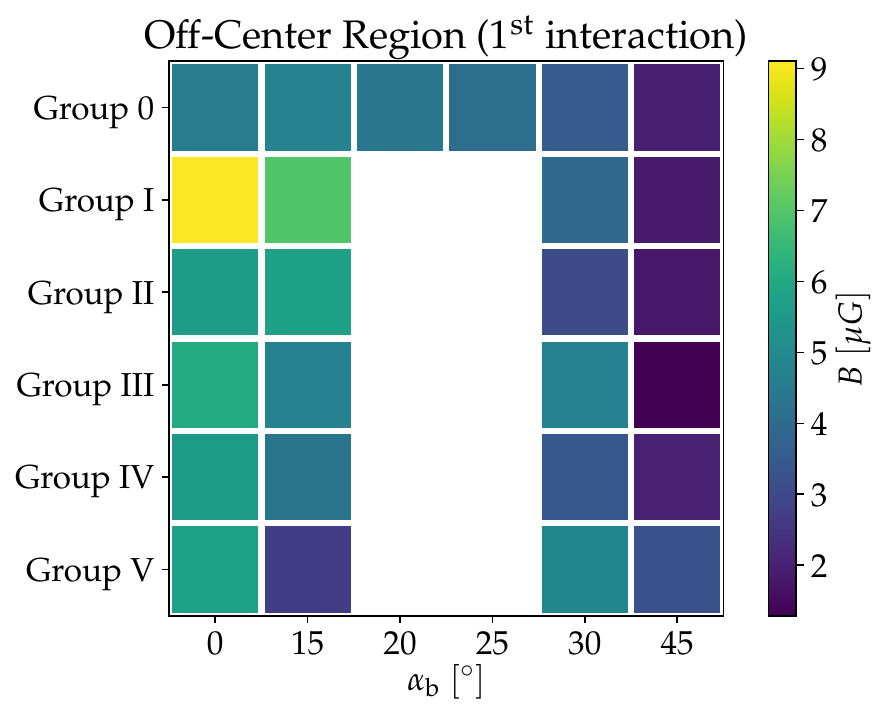}
    \includegraphics[width=0.45\textwidth]{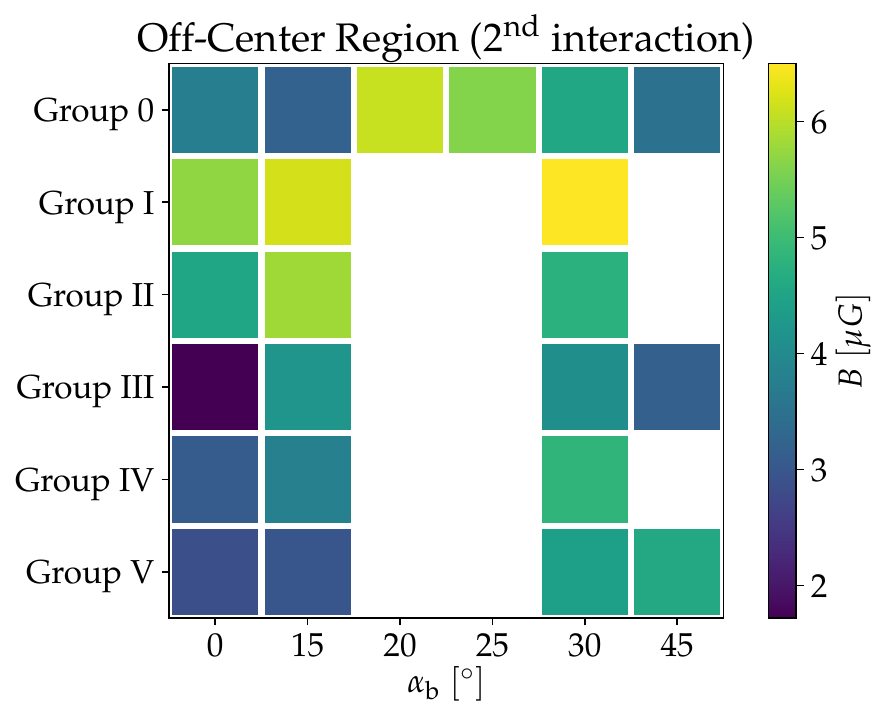}
    \caption{Categorical heat maps displaying the magnetic field strength at the first interaction, ordered by impact parameter angle ($\alpha_{\mathrm{b}}$) and simulation group. Simulations are grouped by common initial inclinations of galaxies, i.e., analogous to Table~\ref{tb:parameters} Group 0: 1-6, Group I: No. 7-10, Group II: 11-14, Group III: 15-18, Group IV: 19-22, Group V: 23-26. Shown are results for center (\emph{top}) and off-center (\emph{bottom}) regions at the first (\emph{left}) and second (\emph{right}) encounters. Parameter combinations that are not covered by our simulation suite or simulations that were not evolved over a sufficiently long time to reach the second interaction are left white.}
    \label{fig:heatmaps}
\end{figure*}

For a systematic comparison of peak heights, we created categorical heat maps of the maximal average magnetic field during first and second encounters (see Fig.~\ref{fig:heatmaps}). In horizontal direction, one can see that the field strength tends to decrease with increasing impact parameter during the first encounters of the disks, while second encounters produce the high peaks also in off-center collisions. This is expected because the separation between the disks in an off-center collision becomes sufficiently small only during the second or third encounter (see also Fig.~\ref{fig:separation}). In contrast, the second peak of the central ($\alpha_\mathrm{b}=0^\circ$) edge-on (group 0) collision is significantly lower than first peak in the central region. Among the central collisions in all groups (vertical direction in the heat maps), the first interactions in group I (relative inclination of $90^\circ$) stands out both in center and off-center regions. The central collision in group II (relative inclination of $90^\circ$) exhibits strong amplification in the center regions during the second encounter (Fig.~\ref{fig:heatmaps} upper right). In these cases, the proximity of tilted disks creates particularly strong tidal forces. If the disks are aligned, on the other hand, they will largely overlap at the closest approach, resulting in smaller differential forces. For $\alpha_\mathrm{b}\ge 15^\circ$ (second columns), one can discern relatively low average field strengths in groups III to V compared to groups 0 to II ($i_1=90^\circ$). This highlights the results shown in Fig.~\ref{fig:FieldEvol}. Ignoring outliers for specific disk regions, the overall field amplification appears to be reduced if both disks are inclined with respect to their relative motion. In the case of face-on collisions (group III), the phenomenological discussion in Sect.~\ref{sec:phenomenology} suggests that perturbations in the disk are less efficient in amplifying the magnetic field. Groups IV and V are intermediate cases between edge-on and face-on collisions. In these groups, relatively high peaks are found in collisions with large impact parameters ($\alpha_\mathrm{b}=30^\circ$ or $45^\circ$) during the second encounter.

In virtually all cases, the peaks in the time evolution of the magnetic field strength are narrow (see Fig.~\ref{fig:FieldEvol}), indicating that magnetic field amplification induced by interactions is transient. This behavior can be explained by competing processes. As can be seen in Fig.~\ref{fig:collision}, the magnetic field spreads over surrounding regions through tidal interactions, implying a gradual loss of magnetic field energy inside the disks. Field amplification by shock compression and turbulence acts against these losses. However, these processes are relatively short-lived. 


\subsection{Mean-field analysis}

To analyze dynamo activity in our simulations, we decomposed the numerically computed velocity and magnetic field into mean-field and turbulent components:
\begin{align}
    \vec{B} &=\overline{\vec{B}} + \vec{B}_{\mathrm{turb}}\,,\\ 
    \vec{v} &=\overline{\vec{v}} + \vec{v}_{\mathrm{turb}}\,. 
\end{align}
We followed the approach of \citet{Ntormousi2020} and applied a mass-averaged filter with smoothing scale $\overline{\Delta}$ in every spatial dimension to compute the mean-field components. We chose $\overline{\Delta} = 300\;$pc, which corresponds to an average over $5^3$ grid cells (five-point stencil in one dimension). The resulting fluctuations, which are given by the difference between the local and mean fields, are associated with the smallest numerically resolved scales. While the smoothing scale applied by \citet{Ntormousi2020} is comparable to the scale height of a thin cooling disk and supernova-driven bubbles in the disk, our choice can be regarded as a compromise between constraining fluctuations to scales comparable to the grid resolution scale and the strong damping of numerical dissipation in this range of scales (see also \citealt{Grete2017}).

As shown in \citet{Brandenburg05}, section 6.2, the resulting mean-field induction equation reads 
\begin{equation}
    \label{eq:induction}
    \frac{\partial\overline{\vec{B}}}{\partial t} = 
    \boldsymbol{\nabla}\times
    \left(\overline{\vec{v}}\times\overline{\vec{B}} + \mathcal{E}\right)
\end{equation}
if the magnetic diffusivity $\eta$ is neglected (ideal MHD approximation). The small-scale electromotive force (EMF) is defined by
\begin{equation}
    \mathcal{E} = \overline{\vec{v}_{\mathrm{turb}}\times\vec{B}_{\mathrm{turb}}}\,. \label{eq:emf}
\end{equation}
In mean-field theories, the small-scale EMF is expanded in powers of the mean magnetic field and current density and closures are applied for the transport coefficients \citep[see][section 6.3]{Brandenburg05}. Here, we explicitly evaluate the right-hand side of Eq.~(\ref{eq:emf}) from the differences between filtered and local fields. This implies that the value of the EMF depends on the numerical resolution. However, for a fixed resolution, we can compare the evolution of the EMF among different interaction scenarios.

\begin{figure*}
    \centering
    \includegraphics[width=0.49\textwidth]{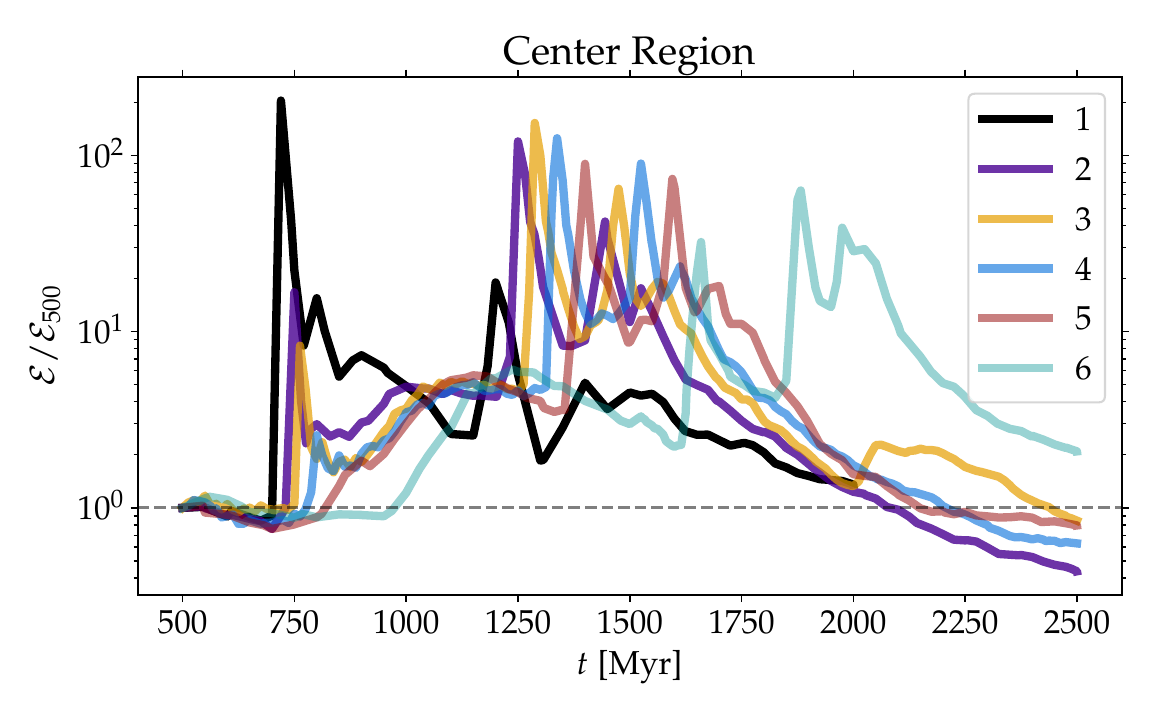}
    \includegraphics[width=0.49\textwidth]{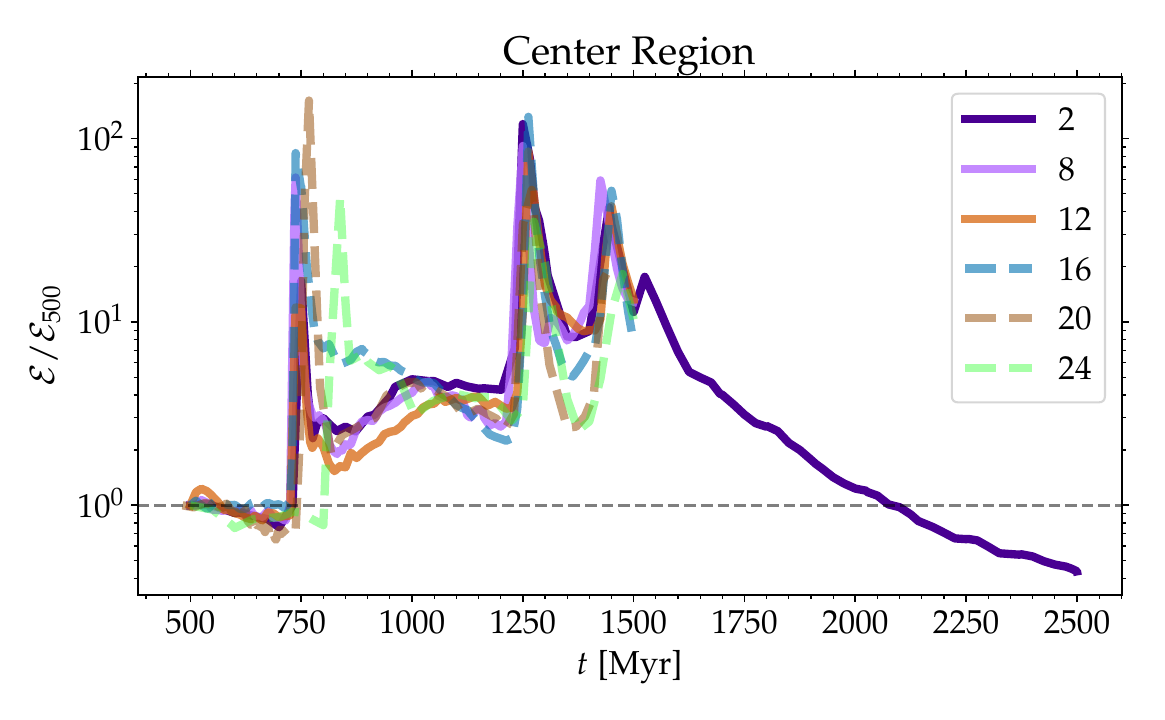} \\
    \includegraphics[width=0.49\textwidth]{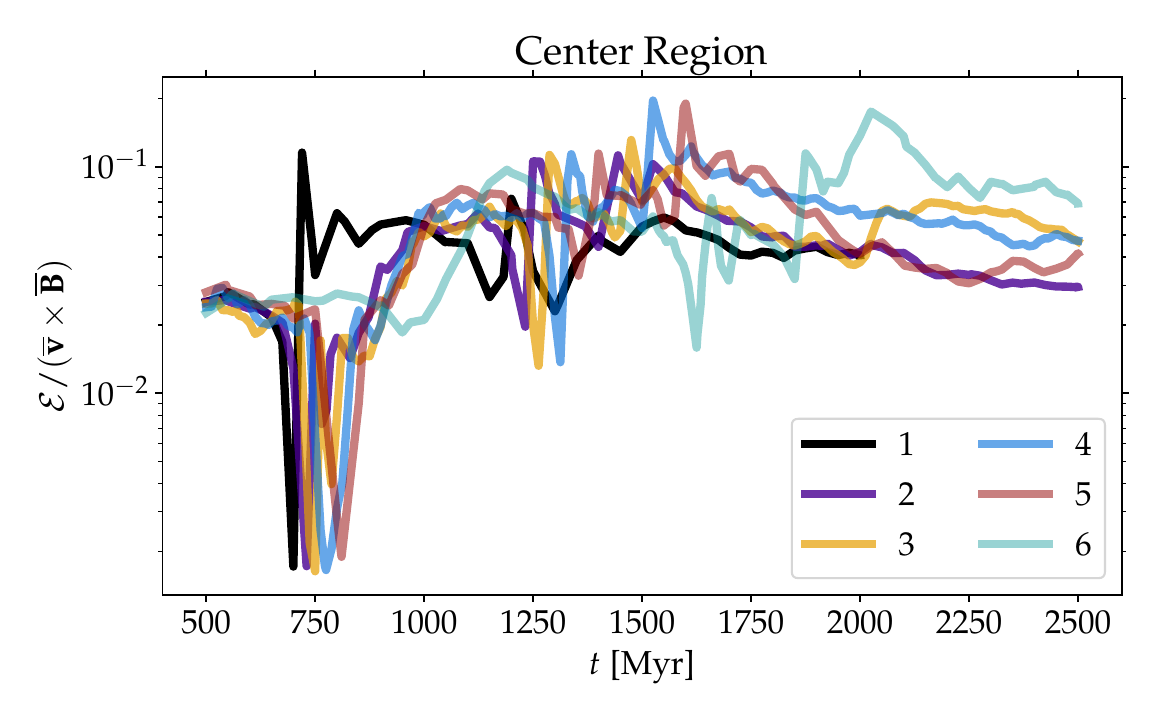}
    \includegraphics[width=0.49\textwidth]{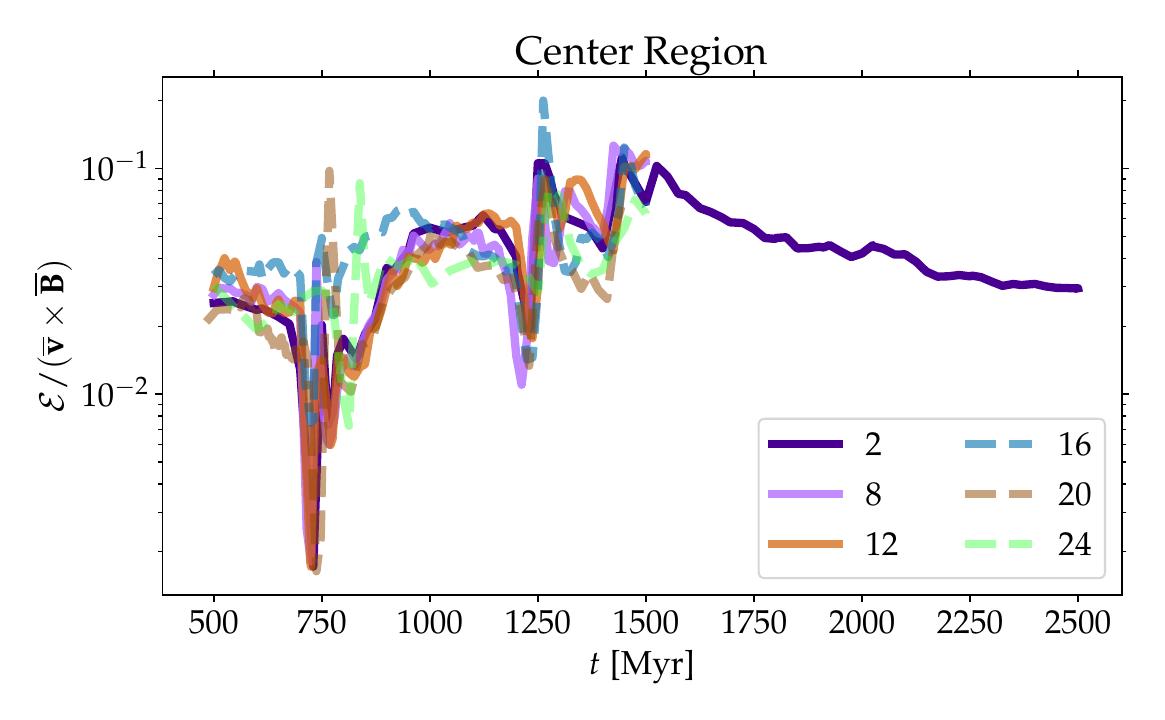}
    \caption{Small-scale EMF, $\mathcal{E,}$ averaged over center regions for simulations of edge-on collisions (\emph{left}) and simulations with the impact parameter $\alpha_{\mathrm{b}}=15^{\circ}$ (\emph{right}). \emph{Top row}: EMF normalized with respect to its value at $t=500\;$Myr. \emph{Bottom row}: $\mathcal{E}$ relative to $\overline{\mathbf{v}}\times\overline{\mathbf{B}}$ (see Eq.~\ref{eq:induction}) averaged over center regions for the same  simulations as in the top row. The corresponding magnetic field evolution is shown in Fig.~\ref{fig:FieldEvol}. }
    \label{fig:EMFCenter}
\end{figure*}

Figure~\ref{fig:EMFCenter} shows the evolution of the small-scale EMF averaged over center regions for the same simulations as in Fig.~\ref{fig:FieldEvol}. In each case, we normalized $\mathcal{E}$ by $\mathcal{E}_{500}$ (i.e.,~the average vale at $t=500\;$Myr). At that point, the galaxies are still in the phase of the first approach, but perturbations in the initial disks have largely relaxed. Thereby, changes in $\mathcal{E}$ with respect to the state of the galaxies before the first encounter when $\mathcal{E}/\mathcal{E}_{500}\approx 1$ can be readily seen. Moreover, we concentrate the discussion on center regions. There is a steep rise of $\mathcal{E}$ in at the first encounter in nearly central collisions. In this phase, $\mathcal{E}/\mathcal{E}_{500}$ increases by up to two orders of magnitude. Only for larger impact parameters, the increase is smaller or even delayed until the second encounter (\simno~5 and~6). In most simulations, the maximum small-scale EMF is reached around the second encounters. An exception is the central edge-on collisions (\simno~1), where by far the strongest amplification occurs during the initial interaction. Apart from that, small-scale perturbations and, thus, turbulence tend to grow in subsequent interactions. However, this evolution is not sustained. In every simulation that was advanced for a sufficiently long period of time, we see that $\mathcal{E}$ gradually decreases in the final phase, when the galaxies have merged and small-scale flow is decaying. Although the averaged magnetic field is 
produced by both the mean-field and the small-scale EMF, its time evolution (Fig.~\ref{fig:FieldEvol}, upper left and right) roughly reflects the evolution of $\mathcal{E}$. This suggests that $\mathcal{E}$ contributes significantly to the amplification of the magnetic field, particularly during close encounters. In the intermediate phases between encounters, $\mathcal{E}/\mathcal{E}_{500}\sim 10$ in most cases (see Fig.~\ref{fig:EMFCenter}), while the average magnetic field decreases.

To investigate the contribution of the small-scale part $\mathcal{E}$ relative to the mean-field part $\overline{\vec{v}}\times\overline{\vec{B}}$ of the EMF in Eq.~(\ref{eq:induction}), we averaged the ratio of these two contributions over the disk regions. Results for center regions are shown in the bottom plots in Fig.~\ref{fig:EMFCenter}. Generally, first encounters are associated with downward spikes, indicating that magnetic field amplification is mostly driven by large-scale processes, such as compression and tidal stresses. However, the ratio $\mathcal{E}/(\overline{\vec{v}}\times\overline{\vec{B}})$ rises in the aftermath of the initial interaction to levels of about $5$ to $10\,\%$. This can be understood as a consequence of turbulence production, as illustrated in Fig.~\ref{fig:LicPostInteraction}. For fixed impact parameter (Figure~\ref{fig:EMFCenter}, lower right), one can clearly see that the ratio also drops at the second encounter, but the small-scale EMF continues to grow until coalescence. In the post-merger phase, both contributions gradually decay. In the group 0 (Fig.~\ref{fig:EMFCenter}, lower left), the pre-interaction ratio of about $0.03$ is reached after about $2.5\mathrm{Gyr}$ for nearly central collisions. For larger impact parameters, the small-scale EMF tends to persists longer (\simno~4 to 6). This indicates that turbulence is mainly produced during late stages, when the mean-field contribution is already relatively small. 

\begin{figure*}
    \centering
    \includegraphics[width=0.49\textwidth]{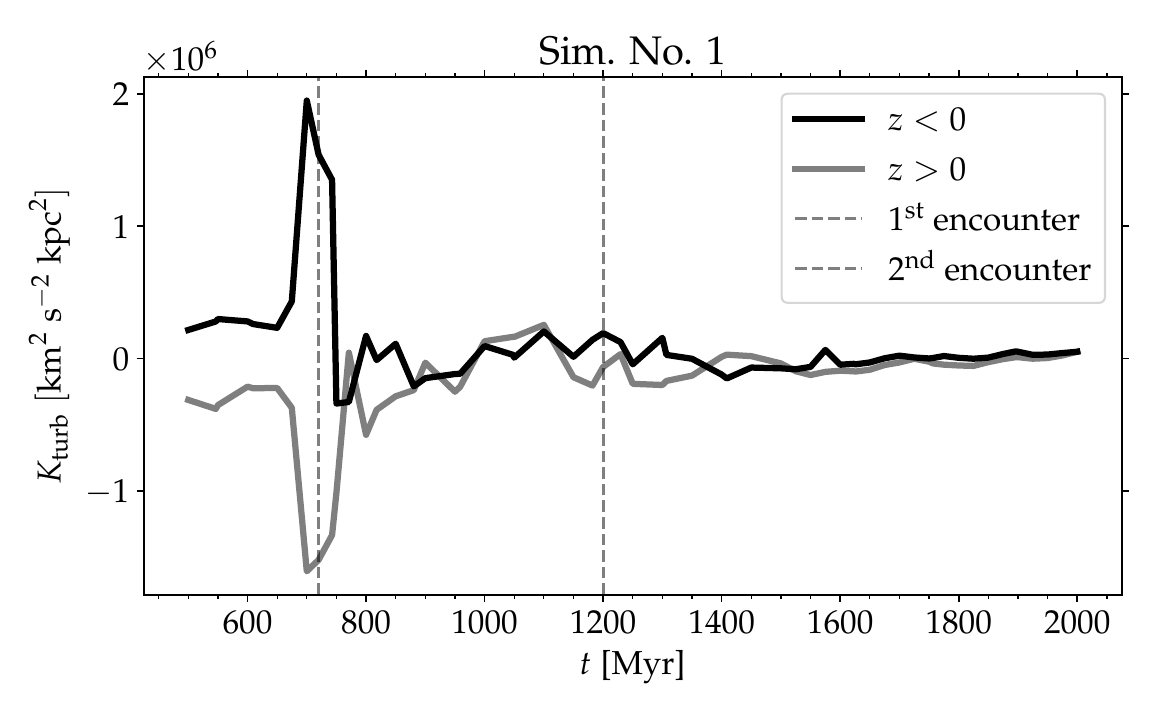}
    \includegraphics[width=0.49\textwidth]{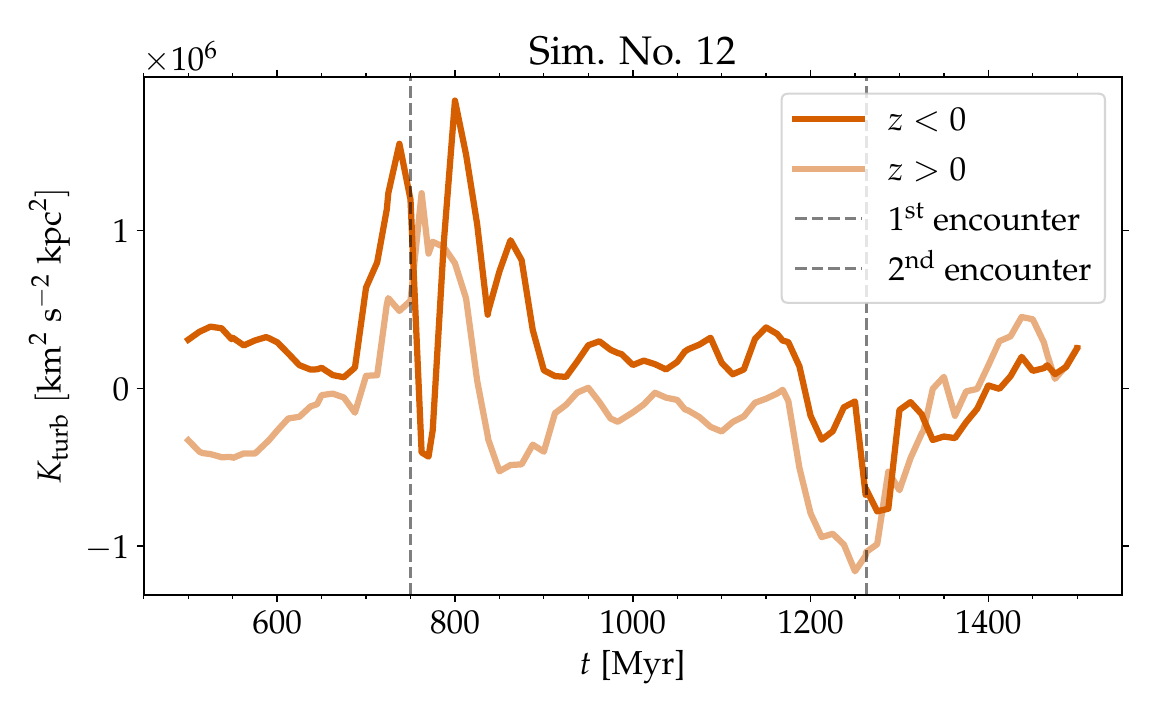} \\
    \includegraphics[width=0.49\textwidth]{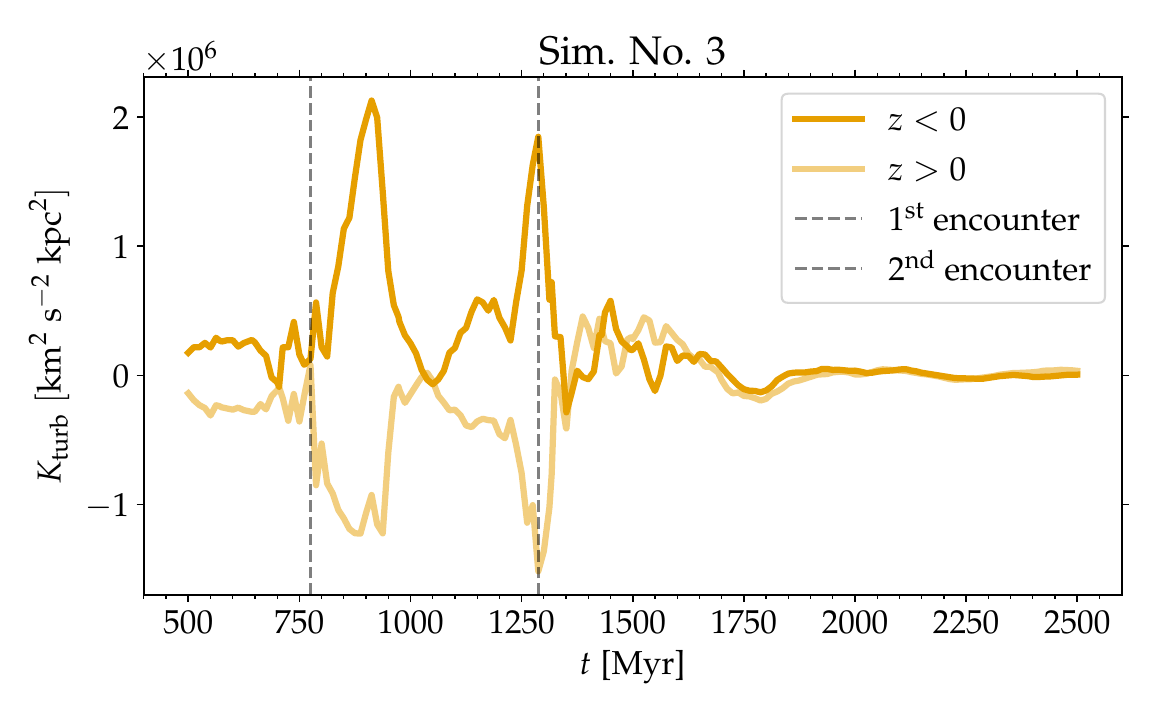}
    \includegraphics[width=0.49\textwidth]{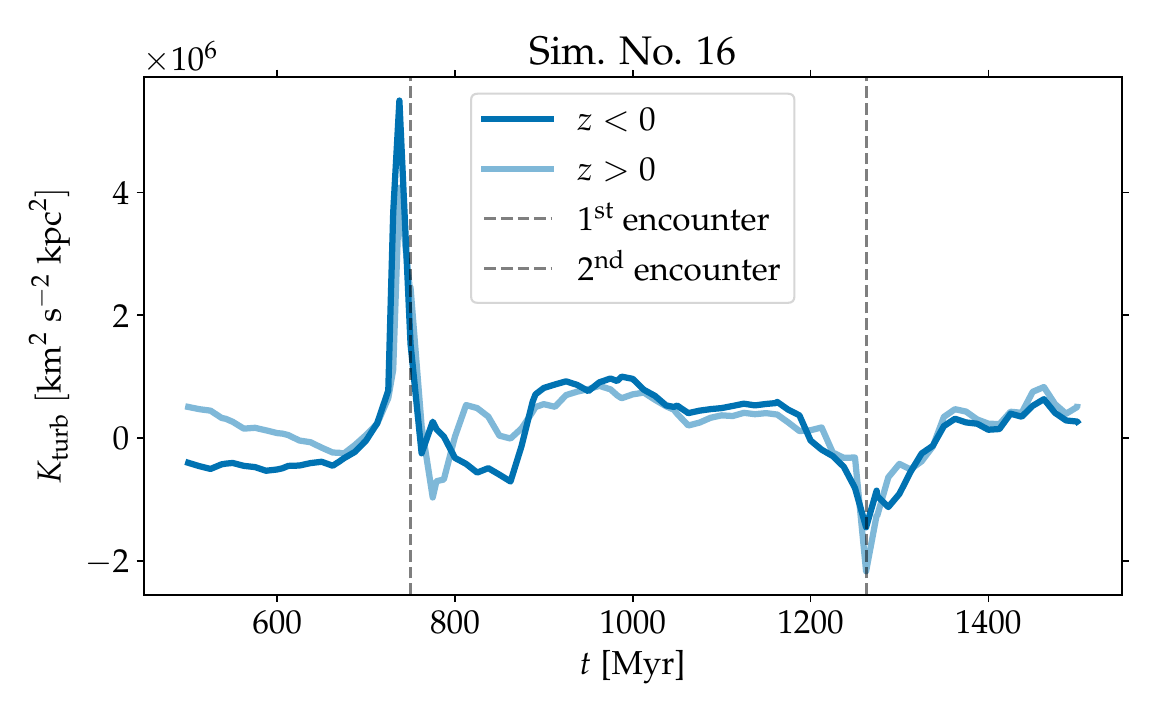} \\
    \includegraphics[width=0.49\textwidth]{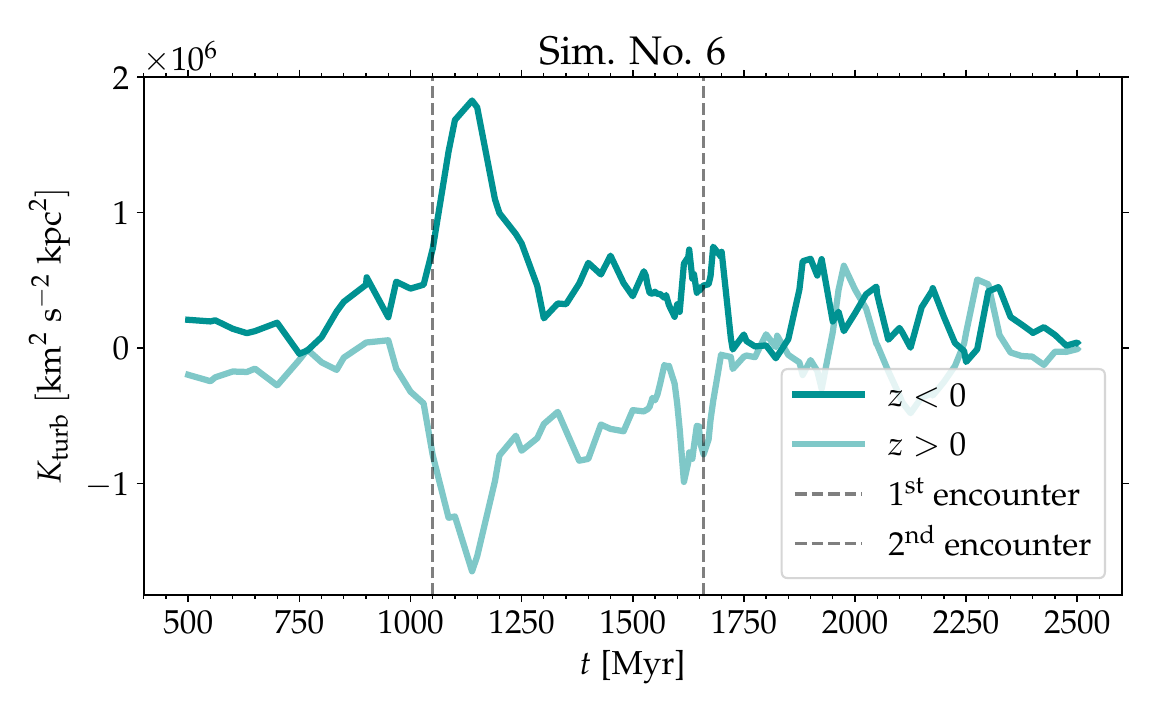}
    \includegraphics[width=0.49\textwidth]{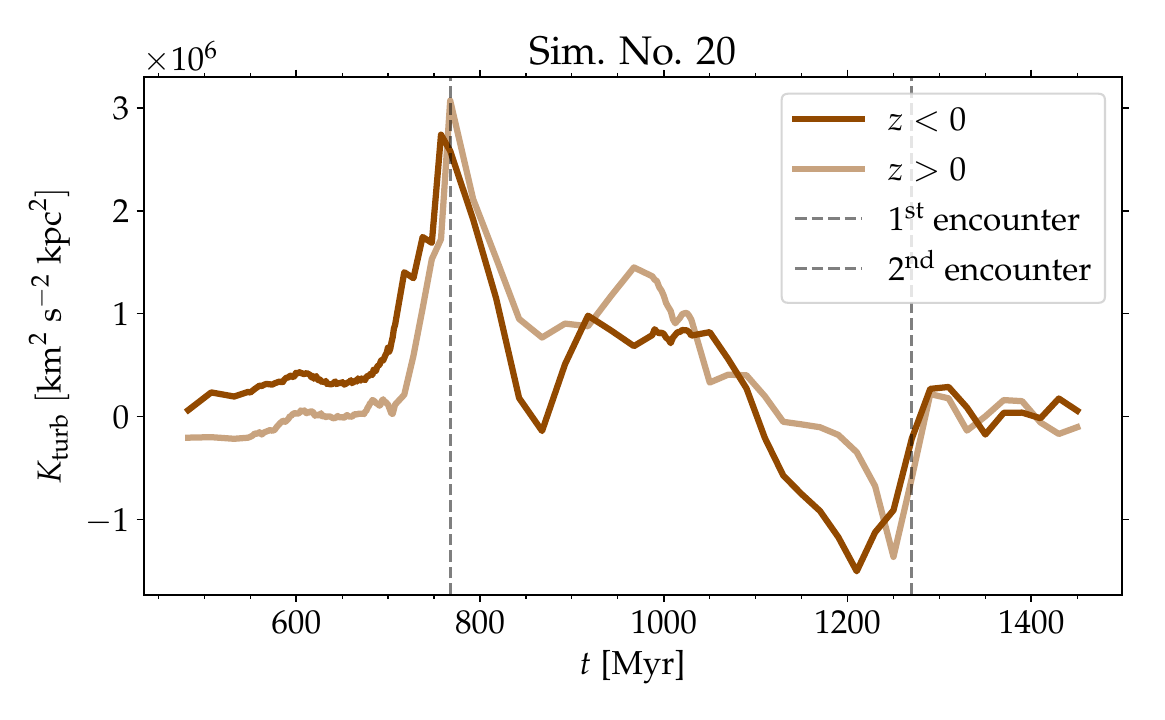}
    \caption{Turbulent kinetic helicity defined by Eq.~(\ref{eq:induction}) averaged over centered cylinders ($r\le 5\;\mathrm{kpc}$, $|z|\le 4\;\mathrm{kpc}$) below and above the galaxy disk midplane. \emph{Left column}: Simulations of edge-on collisions (group 0 with $\alpha_{\mathrm{b}}=0^{\circ}$, $20^{\circ}$, and $45^{\circ}$). \emph{Right column}: Simulations with the impact parameter $\alpha_{\mathrm{b}}=15^{\circ}$ (groups II, III, and IV).}
    \label{fig:TurbKinHel}
\end{figure*}

One of the simplest dynamo models, the so-called $\alpha$ effect, assumes that $\mathcal{E}$ is related to the turbulent kinetic helicity \citep[see][section 6.3]{Brandenburg05}. The turbulent kinetic helicity is defined by the integral
\begin{equation}
    K_{\mathrm{turb}} = \int{\vec{v}_\mathrm{turb}\cdot(\boldsymbol{\nabla}\times\vec{v}_\mathrm{turb})\;\mathrm{d}V}\,,
\end{equation}
which we evaluated within two adjacent cylindrical volumes that are immediately below ($K_{\mathrm{turb}}^<$) and above ($K_{\mathrm{turb}}^>$) the midplane, respectively \citep[see also][]{Ntormousi2020}. With a thickness of $4\;$kpc and a radial extent of $20\;$kpc, the combined volume above and below the midplane encompasses the center and off-center regions for computing averages of the  magnetic field strength. Since helicity is a pseudoscalar, reflection across the midplane reverses its sign \citep{Yokoi2013}. Prior to interaction, $K_{\mathrm{turb}}^>$ and $K_{\mathrm{turb}}^<$ are comparable and have opposite signs. The time evolution is shown for selected simulations in Fig.~\ref{fig:TurbKinHel}. In the case of a central edge-on collision (\simno~0), $K_{\mathrm{turb}}^>$ has a pronounced negative peak at the first encounter, which is roughly balanced by a positive peak of $K_{\mathrm{turb}}^<$. This phase is dominated by strong helical flows inside the tidal arms and the tidal bridge connecting the retracting collision partners. Assuming a linear relation between $\mathcal{E}$ and $K_{\mathrm{turb}}$ and an initially coherent toroidal mean field, the $\alpha$ effect preferentially amplifies the magnetic field in opposite directions above and below the midplane, which results in a quadrupolar field component \citep{Ntormousi2020}. After the first encounter, both  $K_{\mathrm{turb}}^>$ and $K_{\mathrm{turb}}^<$ become small and irregular. For larger impact parameters (Fig.~\ref{fig:TurbKinHel}, \simno~3 and 6), $K_{\mathrm{turb}}^>$ and $K_{\mathrm{turb}}^<$ remain antisymmetric in the post-interaction phase and throughout the second encounter. In the case of \simno~3 ($\alpha_{\mathrm{b}}=20^{\circ}$), antisymmetry is eventually broken in the coalescence phase. A similar evolution is seen in \simno~6 ($\alpha_{\mathrm{b}}=45^{\circ}$), but instead of the pronounced peaks at the second encounter, the average kinetic helicity decreases gradually at late time. Comparing to Fig.~\ref{fig:EMFCenter}, pronounced antisymmetric peaks of $K_{\mathrm{turb}}^>$ and $K_{\mathrm{turb}}^<$ are associated with a decrease in $\mathcal{E}/(\overline{\vec{v}}\times\overline{\vec{B}})$. If the magnetic field is still more or less toroidal at that point, the resulting small-scale EMF tends to be antisymmetric and cancels out over the whole cylindrical region. The increasing EMF in post-interaction phases, on the other hand, indicates a growing influence of non-helical isotropic turbulence.

For scenarios where the disk inclination differs from $90^\circ$, we see a markedly different evolution (Fig.~\ref{fig:TurbKinHel}, right column). As an example for face-on collisions (group III), \simno~16 is shown. In this case, $K_{\mathrm{turb}}^>$ and $K_{\mathrm{turb}}^<$ have the same sign after the initial approach, meaning that the contributions from above and below the midplane are symmetric rather than antisymmetric. 
Since the two disks have parallel angular momentum, one would expect that the mean kinetic helicity of the lower half of one disk cancels the upper half of the other disk when they overlap. However, it appears that tidal interactions between the two disks fundamentally change their structure, resulting in a positive (first encounter) or negative (second encounter) net helicity. To a lesser degree this can also be seen in \simno~20 (group IV, both disks are inclined), while \simno~12  (group II, only secondary disk has inclination $0^\circ$) is a mixed case with alternating antisymmetric and symmetric episodes. These qualitative differences in the kinetic helicity evolution correspond well to differences in the small-scale EMF and the relatively weak magnetic field amplification in groups III to V, where both disks are inclined.


\section{Conclusions}

We explored the evolution of galactic magnetic fields during interactions and mergers of disk galaxies by means of numerical simulations. Gaseous disks in hydrostatic equilibrium were initialized using the potential method from \cite{Wang10}. Moreover, we adopted the method from \citet{Drakos17} to compute particle initial conditions for live dark-matter halos. To compute the evolution of disks and halos, we employed the MHD AMR code Enzo. This allowed us to analyze magnetic field amplification induced by the tidal interactions between the galaxies. In an elaborate parameter study, we varied the impact parameter (i.e., the normal distance of the trajectory of the inbound galaxy to the target galaxy) and the relative orientations of disks.

During interaction phases, we find pronounced peaks in the magnetic field strength averaged over the center and outer regions of the disks. These peaks are a distinctive feature that was also identified by \citet{Drzazga11} in radio observations of interacting galaxies. However, there are important differences. For all the galaxies in their sample, \citet{Drzazga11} found a relatively small increase during encounters, and the strongest fields they found are associated with the final coalescence phase. In our suite of simulations, this is only the case for impact parameters $\alpha_\mathrm{b}$ (the angle defined by the ratio of the linear impact parameter and the initial separation) greater than about $20^\circ$ or, to a lesser degree, for collisions of disks that are inclined with respect to their orbital motion and whose axes are not aligned (Fig.~\ref{fig:FieldEvol}). Nearly central collisions typically exhibit the highest peaks during the first encounter. If these trends are correct, this implies that nearly central collisions are rare and were not present in the small galaxy sample investigated by \citet{Drzazga11}, who ordered the galaxies by interaction stages. However, our results indicate that, even at the same interaction stage, variations in the field strengths may result from different interaction parameters. Since three-dimensional data are required, it will be challenging to lift this degeneracy by classifying observed systems accordingly.

Moreover, the mean field strength in center regions exceeds $10\;\upmu$G in about half of the scenarios we investigated, while the maximum mean field strength is typically in the range from $5$ to $10\;\upmu$G in off-center regions (Fig.~\ref{fig:heatmaps}). These values are systematically lower than the field strengths inferred by \citet{Drzazga11} from nonthermal radio emission, which reach $50\;\upmu$G and $20\;\upmu$G in center and off-center regions, respectively. This discrepancy might be a consequence of insufficient numerical resolution resulting in a less effective turbulent dynamo. However, comparable field strengths were obtained by \citet{Rodenbeck2016} despite the significantly lower resolution of their simulations ($240\;$pc compared to $15\;$pc in our simulations). The most likely explanation is the assumption of adiabatic gas dynamics. Since the gas cannot cool, adiabatic disks do not contract into thin disks in our simulations. Thus, they have larger scale heights and the initial magnetic field undergoes less compression (assuming flux conservation), resulting in an overall lower field strength. This is supported by relative amplification factors of about $2$ to $3$ (Fig.~\ref{fig:FieldEvol}), which compares well to the \citet{Drzazga11} results. In addition, we neglected magnetic field amplification due to turbulence driven by feedback processes. \citet{Whittingham21} do reproduce observed galactic field strengths in cosmological zoom-in simulations with nonadiabatic physics and stellar feedback.

As a quantitative indicator, \cite{Whittingham21} show that the power spectra of the magnetic field energy in the central regions of interacting galaxies approximately agree with Kolmogorov scaling toward small length scales. This suggests that the field is turbulent \citep{beresnyak_mhd_2019}. However, power spectra are of limited validity for complex structures such as interacting galaxies, where the intensity of turbulence undergoes substantial spatial variations and rapid changes in time. Moreover, there are localized compression effects, for example shock compression of gas in collision zones. For this reason, we applied a box filter to decompose the velocity and magnetic field into a mean-field part and small-scale fluctuations.

The amplification of the magnetic field during encounters and coalescence is reflected by an enhanced EMF, which subsides in the aftermath of the interaction (Fig.~\ref{fig:EMFCenter}, top row). As a result, the amplification is transient. Basically, this agrees with the observational data from \citet{Drzazga11} and the numerical studies \citep{Pakmor2014,Rodenbeck2016,Whittingham21}. By comparing small-scale and mean-field components, we find that the turbulent, small-scale EMF grows after the first interaction to a fraction on the order of $0.1$ of the mean-field contribution. However, when the galaxies approach each other and get close to the orbital pericenter, the ratio of small-scale to mean-field EMF decreases in many cases (Fig.~\ref{fig:EMFCenter}, bottom row). These minima are typically associated with antisymmetric peaks of the turbulent kinetic helicity above and below the midplane (Fig.~\ref{fig:TurbKinHel}). Antisymmetric kinetic helicity was previously reported by \citet{Ntormousi2020} for isolated disk galaxies. As pointed out by \citet{Brandenburg05}, inhomogeneous and anisotropic flows typically have large helicities and act as large-scale dynamos. Such flows are produced by tidal forces, and, consequently, the mean-field contribution should be relatively large during close encounters. The small-scale EMF becomes more significant in post-interaction phases in which the gas becomes increasingly turbulent and the galaxies finally merge. The turbulent kinetic helicity switches from antisymmetry to an irregular regime and decays in the remnant. If the rotation axis of both galaxies are inclined, such as in face-on collisions, a significant net helicity develops during the first encounter. In these scenarios, the maximal amplification of the magnetic field is relatively low (Fig.~\ref{fig:heatmaps}). 

Overall, this led to our interpretation that the peaks in the average magnetic field are mainly due to tidally induced helical flows. This process is more efficient if the disk plane is oriented parallel to relative motion. The production of turbulence enhances the magnetic field, especially in post-interaction phases. However, the small-scale dynamo is not sustained for long after coalescence. A lasting effect is the unordered, random structure of the magnetic field in the remnants, which are always ellipsoids. However, in some cases this shape might be an artifact of adiabatic gas dynamics. It is conceivable that radiative cooling allows for disk-like remnants, where a coherent field could be rejuvenated by the $\alpha$-$\Omega$ dynamo.
Apart from the highly dynamic evolution of magnetic fields, star formation and, in turn, feedback are triggered by galaxy collisions \citep{Moreno21,Whittingham21,Linden22,Renaud22}. Moreover, the dependence of the small-scale dynamo on numerical resolution in galaxy simulations is a pending problem \citep{Grete2019,Koertgen19,Ntormousi2020,Whittingham21,Liu22,Wissing2023}. As shown by \citet{Schober2013}, the small-scale dynamo first amplifies the magnetic field to saturation on the viscous scale. However, this scale is in the sub-parsec range, far below scales that can be resolved in galaxy simulations. Subsequently, the magnetic energy is shifted to larger scales until it saturates on the forcing scale. In a numerical simulation, the smallest scale is set by numerical diffusion. Consequently, the turbulent EMF computed from our simulations initially operates on scales that are orders of magnitude larger than the physical amplification scales. Numerical diffusion also determines the effective resistive scale on which magnetic energy is dissipated. Consequently, growth and decay rates may very well change with resolution. Since we have demonstrated that the small-scale EMF plays a significant role in the turbulent environment of interacting galaxies, improved modeling of the underlying dynamo and dissipation processes will be an important next step.

\begin{acknowledgements}
      We thank Kai Rodenbeck and Dominik Schleicher for their pioneering work and for providing the code for the equilibrium disk setup in Enzo. The work presented in this article was supported by the German \emph{Deut\-sche For\-schungs\-ge\-mein\-schaft, DFG\/} project
      number SCHM~2135/6-1. The simulations were carried out using computational resources granted by HLRN project hhp00049.
\end{acknowledgements}

\bibliographystyle{aa} 
\bibliography{references}

\end{document}